\documentclass[a4paper,fleqn,usenatbib]{mnras}

\usepackage{newtxtext,newtxmath}

\usepackage[T1]{fontenc}
\usepackage{ae,aecompl}

\usepackage{graphicx}	
\usepackage{amsmath}	
\usepackage{amssymb}	

\usepackage{threeparttable} 
\usepackage{pdflscape} 
\usepackage{longtable} 

\usepackage[normalem]{ulem}

\usepackage[dvipsnames]{xcolor}
\newcommand{\rknew}[1]{#1}

\defcitealias{Kokotanekova2017}{K17}

\title[Spin Changes of Cometary Nuclei]{Implications of the Small Spin Changes Measured for Large Jupiter-Family Comet Nuclei}

\author[R. Kokotanekova et al.]{R. Kokotanekova$^{1,2}$\thanks{E-mail: kokotanekova@mps.mpg.de}, C. Snodgrass$^{2}$, P. Lacerda$^{3}$, S. F. Green$^{2}$, P. Nikolov$^{4}$, 
\newauthor T. Bonev$^{4}$
\\
$^{1}$Max Planck Institute for Solar System Research,
              Justus-von-Liebig-Weg 3, 37077, G\"ottingen, Germany\\
$^{2}$Planetary and Space Sciences, School of Physical Sciences, The Open University, Milton Keynes, MK7 6AA, UK\\
$^{3}$Astrophysics Research Centre, Queen's University Belfast, Belfast BT7 1NN, UK\\
$^{4}$Institute of Astronomy and National Astronomical Observatory, 72 Tsarigradsko Chauss\'ee Boulevard, BG-1784 Sofia, Bulgaria \\
}

\date{Accepted XXX. Received YYY; in original form ZZZ}

\pubyear{2018}

\begin{document}
\label{firstpage}
\pagerange{\pageref{firstpage}--\pageref{lastpage}}
\maketitle


\begin{abstract}

Rotational spin-up due to outgassing of comet nuclei has been identified as a possible mechanism for considerable mass-loss and splitting. We report a search for spin changes for three large Jupiter-family comets (JFCs): 14P/Wolf, 143P/Kowal-Mrkos, and 162P/Siding Spring. None of the three comets has detectable period changes, and we set conservative upper limits of 4.2 (14P), 6.6 (143P) and 25 (162P) minutes per orbit. Comparing these results with all eight other JFCs with measured rotational changes, we deduce that none of the observed large JFCs experiences significant spin changes. This suggests that large comet nuclei \rknew{are less likely to undergo rotationally-driven splitting}, and therefore more likely to survive more perihelion passages than smaller nuclei. We find supporting evidence for this hypothesis in the cumulative size distributions of JFCs and dormant comets, as well as in recent numerical studies of cometary orbital dynamics. We added 143P to the sample of 13 other JFCs with known albedos and phase-function slopes. This sample shows a possible correlation of increasing phase-function slopes for larger geometric albedos. Partly based on findings from recent space missions to JFCs, we hypothesise that this correlation corresponds to an evolutionary trend for JFCs. We propose that newly activated JFCs have larger albedos and steeper phase functions, which gradually decrease due to sublimation-driven erosion. If confirmed, this could be used to analyse surface erosion from ground and to distinguish between dormant comets and asteroids.

\end{abstract}





\section{Introduction}
\label{sec:Intro}


It is widely accepted that comets are among the most unaltered bodies in the Solar System. However, they are also known to undergo dramatic changes driven by sublimation activity. Understanding the effects of cometary evolution is therefore key for discerning their primordial properties and relating them to the early Solar-system history.

Having orbital periods of less than twenty years, Jupiter-family comets (JFCs) allow repeated observations over multiple apparitions (and perihelion passages). These observations can be used to monitor the changes in activity, rotation and surface properties experienced by the comets. Moreover, the relatively low eccentricity and inclination of JFCs as well as their relative proximity to Earth has made them accessible to several space missions which have improved the understanding of cometary physics tremendously over the past few decades. 

It is well-established that JFCs were formed beyond the snowline in the early Solar System about 4.6 Gyr ago \citep[see][and references therein]{Davidsson2016}. According to the Nice model \citep{Tsiganis2005,LEVISON2008}, planetary migration of Jupiter and Saturn destabilised the outer Solar System about 400 Myr after the formation of the primordial disk, and scattered the icy planetesimals to form the Kuiper Belt and the scattered disk. These two regions are considered to be the most likely reservoirs of today's JFCs \citep{Levison1997,Duncan1997a}. In other words, after spending about 4 Gyr beyond the orbits of Neptune, some Trans-neptunian objects get destabilised due to interactions with the outer giant planets, and make a return to the inner Solar System as Centaurs and eventually as JFCs. Once the returning small bodies reach heliocentric distances less than 3-5 au, they become active comets characterised by sublimation of water and other volatiles. 

There are a few different scenarios describing the final fates of comets. Most nuclei are believed to either gradually lose their activity until they become dormant or dead comets, or, alternatively, to experience catastrophic comet-splitting events \citep[see][]{Jewitt2004a,Boehnhardt2004}. One of the possible mechanisms leading to comet splitting is activity-driven spin-up. This mechanism takes place when outgassing produces torques which bring the rotation periods of the nuclei down to a critical limit. Below this limit, the centrifugal force exceeds the gravity and the material forces, and the comet nucleus starts to shed mass and falls apart \cite[e.g.][]{Davidsson1999,Davidsson2001}. 

So far, the rotation rates of 37 comets have been determined \citep[see][hereafter K17, and references therein]{Kokotanekova2017}. Repeated observations of eight of them have shown clear indications for spin changes on orbital timescales \citep[see][and references therein]{Samarasinha2013,Eisner2017,Bodewits2018}. Moreover, the direct measurements of the rotation changes of comet 67P/Churyumov--Gerasimenko during the Rosetta mission were successfully reproduced by the numerical model of \cite{Keller2015}. This study confirmed the widely accepted hypothesis that the rotation-period changes are controlled by outgassing torques and depend on the shape and orientation of the cometary nuclei \citep{Keller2015}.

Spin changes of outgassing comets can be described by simple theoretical considerations \citep[e.g.][]{Samarasinha2004,Samarasinha2013}. In particular, these models predict that for comets of identical densities, sizes, shapes, activity levels and active-region distributions, the smaller nuclei will experience larger period changes. The rotation changes of small cometary nuclei were also studied by numerical models using realistic shape models and activity distributions \citep{Gutierrez2005}. These authors confirmed that small active nuclei experience typical changes of 0.01--10 hours per orbit. However, to our knowledge, the spin changes of larger nuclei have not been directly modelled in published works. 

If the nuclei do not undergo significant mass-loss and disruption events during the prime of their activity as JFCs, they are expected to gradually decay in activity until they become dormant (nuclei for which the available volatiles are shielded from solar insolation) or dead (totally devolatilised) comets \citep{Weissman1999,Jewitt2004a}. Due to the lack of detectable activity of these objects it is difficult to distinguish dormant/dead comets from asteroids that have been placed on comet-like orbits \citep{Fernandez2001,Fernandez2005a}.

In this work, we aim to understand the changes that active JFCs experience in terms of rotation and surface properties. We present new lightcurve and phase-function observations of three JFCs with previously studied rotation rates, 14P, 143P and 162P. In Section \ref{sec:observations} we summarize the observations and data-analysis procedures used to derive the new lightcurves of the comets. This is followed by Section \ref{sec:results}, where we show the newly obtained lightcurves and the search for period changes. In Section \ref{sec:discussion}, we first present a line of evidence suggesting that large JFCs (with radii $\geq$ 2-3 km) have an enhanced survivability in comparison to smaller nuclei (Section \ref{sec:disc_survivability}). This is followed by a discussion of our hypothesis that geometric albedos and phase functions of JFCs contain information about the erosion level of the nuclei in Section \ref{sec:disc_surface}. Finally, the results and implications of this work are summarised in Section \ref{sec:conclusions}.

\section{Observations and Data Analysis}
\label{sec:observations}

\subsection{Observing Instruments}
\label{sec:obs_instruments}

The observations analysed in this work were performed between January 2016 and March 2017 using three different telescopes (Table \ref{table_all_obs}). SDSS r' filters were used in all observations.


Some of the observations of  143P and 162P were performed with the 2.5-m Isaac Newton Telescope (INT) at the Roque de Los Muchachos observatory on La Palma, Spain. We used the Wide Field Camera (WFC) which is mounted at the primary focus of the INT. WFC consists of a mosaic of four thinned EEV 2048 $\times$ 4096 pixel CCDs.  Each CCD has an effective field of view of 11.5 $\times$ 23 arcmin\textsuperscript{2} and a pixel scale of 0.33 arcsec/pixel. The data for this work were obtained only from CCD 4.

Comets 14P and 143P were observed using the Large Area Imager for Calar Alto (LAICA) installed at the prime focus of the 3.5m telescope of Calar Alto Observatory in Spain. LAICA has a mosaic of 4 CCDs each with 4000 $\times$ 4000 pixels. The total field of view of LAICA is 44.36' $\times$ 44.36' and the pixel scale is 0.225 arcsec/pixel. Throughout the observations we restricted ourselves to using CCD 1. 

Comets 143P and 162P were also observed with the 2-m Ritchey-Chr{\'e}tien Coud{\'e} telescope of the National Astronomical Observatory Rozhen in Bulgaria. We used the VersArray 1300B CCD camera (1340 $\times$ 1300 pixels) which was attached to the two-channel focal reducer FoReRo-2 with resolution of 0.74 arcsec/pixel and a field of view of about 15 arcmin in diameter.

\begin{table*}
\centering
\caption{Summary of all analysed observations.}
\label{table_all_obs}
\begin{tabular}{llllllllc}
\hline
Comet & UT date & $R_{\mathrm{h}}$ {[}au{]}\textsuperscript{a} & $\Delta$ {[}au{]}\textsuperscript{b} & $\alpha$ {[}deg.{]}\textsuperscript{c} & Number & Exposure time [s] & Instrument & Proposal ID \\
\hline
14P   & 2016-07-06 & 3.93\textsuperscript{I}                      & 3.15                                   & 10.57               & 34     & 24x300, 10x240        & CAHA 3.5m/LAICA  & H16-3.5-032 \\
      & 2016-07-07 & 3.93\textsuperscript{I}                      & 3.15                                   & 10.72               & 24     & 23x300, 1x360         & CAHA 3.5m/LAICA  & H16-3.5-032 \\
      & 2016-07-08 & 3.92\textsuperscript{I}                      & 3.15                                   & 10.87               & 33     & 17x300, 11x240, 5x180 & CAHA 3.5m/LAICA  & H16-3.5-032 \\
      & 2016-07-09 & 3.92\textsuperscript{I}                      & 3.16                                   & 11.01               & 25     & 24x240, 1x300         & CAHA 3.5m/LAICA  & H16-3.5-032 \\
      & 2016-07-10 & 3.92\textsuperscript{I}                      & 3.16                                   & 11.15               & 27     & 15x180, 6x150, 6x120  & CAHA 3.5m/LAICA  & H16-3.5-032 \\
143P  & 2016-01-16 & 5.03\textsuperscript{I}                      & 4.38                                   & 9.10                & 29     & 180                   & CAHA 3.5m/LAICA  & F16-3.5-005 \\
      & 2017-02-17 & 3.73\textsuperscript{I}                      & 3.03                                   & 11.91               & 53     & 180                   & INT/WFC          & I/2017A/05  \\
      & 2017-02-18 & 3.73\textsuperscript{I}                      & 3.04                                   & 12.11               & 40     & 180                   & INT/WFC          & I/2017A/05  \\
      & 2017-02-19 & 3.72\textsuperscript{I}                      & 3.05                                   & 12.30               & 22     & 21x180, 1x60          & INT/WFC          & I/2017A/05  \\
      & 2017-02-21 & 3.72\textsuperscript{I}                      & 3.07                                   & 12.66               & 26     & 18x300, 8x200         & INT/WFC          & I/2017A/05  \\
      & 2017-02-26 & 3.70\textsuperscript{I}                      & 3.11                                   & 13.49               & 34     & 300                   & Rozhen 2m/FoReRo & -           \\
      & 2017-02-27 & 3.69\textsuperscript{I}                      & 3.12                                   & 13.65               & 16     & 300                   & Rozhen 2m/FoReRo & -           \\
      & 2017-03-23 & 3.61\textsuperscript{I}                      & 3.37                                   & 15.98               & 15     & 300                   & Rozhen 2m/FoReRo & -           \\
162P  & 2017-02-17 & 4.30\textsuperscript{O}                      & 3.58                                   & 9.88                & 93     & 120                   & INT/WFC          & I/2017A/05  \\
      & 2017-02-18 & 4.31\textsuperscript{O}                      & 3.57                                   & 9.71                & 52     & 120                   & INT/WFC          & I/2017A/05  \\
      & 2017-02-21 & 4.31\textsuperscript{O}                      & 3.55                                   & 9.18                & 79     & 43x120, 36x150        & INT/WFC          & I/2017A/05  \\
      & 2017-02-26 & 4.33\textsuperscript{O}                      & 3.51                                   & 8.24                & 21     & 300                   & Rozhen 2m/FoReRo & -                  \\
 \hline
\end{tabular}
\begin{minipage}{\textwidth}
	\textsuperscript{a} Heliocentric distance. Superscripts I and O indicate whether the comet was inbound (pre-perihelion) or outbound (post-perihelion). \\
    \textsuperscript{b} Geocentric distance. \\
    \textsuperscript{c} Phase angle \\
\end{minipage}
\end{table*}

\subsection{Data reduction and photometry}
\label{sec:data_reduction}

The data analysis techniques used in this paper are explained in detail in \citetalias{Kokotanekova2017}; we summarise them below. Data reduction was done using standard IRAF tasks \citep{Tody1986,Tody1993} from the PyRAF package\footnote{\url{http://www.stsci.edu/institute/software_hardware/pyraf}}. Firstly, a nightly master bias frame was created and subtracted from every frame. Depending on the availability of twilight flats, we median combined either sky or dome flats to create a master flat frame for each night. Finally, each bias-subtracted sky image was divided by the master flat frame.   

The brightness variations of the comets were determined by differential photometry with respect to carefully selected stars common to all frames of the corresponding night. The instrumental magnitudes of the comets, as well as the comparison stars, were measured from aperture photometry using small apertures (typically equal to the FWHM of the PSF on the frame). Since the instruments used in this work have large fields of view, we corrected the instrumental magnitudes for the specific distortions of each instrument, identified as small position-dependent systematics in the aperture photometry of the field stars \cite[see][for INT/WFC]{Hodgkin2008}. To analyse the data taken with FoReRo, we used larger apertures of 1.6 times the FWHM of the PSF to compensate for image distortions.

The comet magnitudes for each night were then calibrated using one reference frame per field (the frame with the best seeing). As in \citetalias{Kokotanekova2017}, absolute photometric calibration using star magnitudes from the Pan-STARRS (PS1) Data release 1 \cite[see][]{Kaiser2002,Kaiser2010,Chambers2016} was performed to convert the instrumental magnitudes of the comets to magnitudes in the Pan-STARRS $\mathrm{r_{P1}}$ system. This was done after the colour term for each of the three instrument configurations was derived following the procedure in \citetalias{Kokotanekova2017}. However, the colour terms for all used instrument configurations were very small and this correction did not have a large effect on the results. Next, we derived a zero point for each reference frame and used it to shift all frames for the corresponding field in order to derive the frame magnitudes $m_\mathrm{r}$.


Finally, all points were corrected for \rknew{light-travel time and solar phase angle effects}. We corrected the magnitudes $m_\mathrm{r}$ for the heliocentric and geocentric distances to obtain $m_{\mathrm{r}}(1,1,\alpha)$ magnitudes. Then the phase-curve effects were removed as part of the Monte Carlo procedure described below, and we finally computed the absolute magnitudes $H_{\mathrm{r}}(1,1,0)$ or $H_{\mathrm{r}}$ in short.

Before combining the data taken at the different observing epochs, we checked whether the comets showed signs of activity during any of the observing runs. This was done following the procedure from \citetalias{Kokotanekova2017}, which compares the average comet PSF profile to that of a neighbouring star. All three comets appeared to have stellar profiles, and we therefore concluded that they were not active during the time of the observations.

\subsection{Monte Carlo method to determine the lightcurve periods}
\label{sec:mc_methods}

In \citetalias{Kokotanekova2017} we used a Monte Carlo method to derive the phase-function slopes and the rotation periods of JFCs from sparsely sampled observations. This technique was chosen because it allowed us to account for the uncertainties occurring at every step of the data analysis: from the differential photometry, from the absolute photometric calibration and from the phase-function correction. It also has the benefit of providing uncertainty ranges of the derived phase-function slopes and periods. However, the downside of this approach is that it uses linear regression to fit a phase function to the data in each of the MC clones. We have confirmed that the linear fitting works very well when the whole range of the lightcurve variation and a broad range of phase angles are sampled. However, in certain cases when the datasets which need to be combined probe the lightcurves just partially, a simple linear fit may produce erroneous results. 

In this work the main goal is to constrain the rotational periods with great accuracy in order to look for spin changes in comparison to previous epochs. To achieve this, we modified our Monte Carlo procedure to consider the entire range of possible phase-function slopes, rather than using only the slopes derived from a linear regression fit to the points in each clone. This has the advantage that a broader range of possible phase-function slopes are tested and therefore the derived possible rotation period range is less dependent on the adopted phase function correction. 


The improved Monte Carlo method (referred to as MC2, hereafter) is based on the MC method used in \citetalias{Kokotanekova2017}. The modified procedure consists of the following steps:

\begin{enumerate}
\item{At each iteration $i$, every magnitude point is replaced by a clone. The clone is a randomly selected value from a normal distribution with standard deviation equal to the photometric uncertainty and mean equal to the original magnitude. }

\item{Next, we shift the clones to account for the uncertainty of the absolute photometric correction. All points belonging to the same calibration star field are shifted with a value randomly selected from a normal distribution with mean equal to 0, and standard deviation equal to the uncertainty of the absolute photometric correction of the given field.}

\item{Then, all points from the produced clone $i$ are corrected for a linear phase function with slope $\beta_i$. The slope is randomly selected from a uniform distribution of phase-function slopes in the range 0.0 to 0.1 mag/deg. To account for the possibility of extreme phase functions, the selected phase-function slopes cover a slightly larger range than the total range of observed phase-function slopes of JFCs \citepalias[0.02-0.08 mag/deg,][]{Kokotanekova2017}.}

\item{To find the best-fitting period $P_{\mathrm{i}}$, we use the \texttt{gatspy}\footnote{\url{http://www.astroml.org/gatspy/}} \texttt{LombScargleFast} implementation \citep{VanderPlas2015} of the Lomb-Scargle method \citep[LS;][]{Lomb1976,Scargle1982}. Experience has shown that the best periods from LS periodograms result in single-peaked lightcurves. Since we assume that the brightness variation of comet nuclei is produced by their shapes, we expect their lightcurves to be double-peaked. Therefore, we double the LS output to get the rotation periods $P_{\mathrm{i}}$. \rknew{The rotation periods determined by this method do not account for changes in the Sun-comet-Earth geometry and are therefore synodic periods. It is impossible to derive the corresponding sidereal periods without information on the polar orientation of the nuclei. However, the difference between the synodic and sidereal rotation periods is expected to be very small when the objects are observed close to opposition \citep{Harris1984}, which is the typical configuration for observing bare comet nuclei.}}


\item{For each clone we phase all points with the period $P_{\mathrm{i}}$ and compute the total string length of the phased lightcurve. The string length is the sum of the distances between the phased magnitude points and follows the definition in the string-length method for period search \citep[SLM;][]{Dworetsky1983}. According to SLM, the lightcurves with shorter total string lengths are more confined and are therefore considered to be better.  }

\item{After repeating this procedure for $i$=1,2,...,5000, we use the distribution of the selected best periods and the corresponding total string lengths for each clone to determine the most likely rotation period and its uncertainty. }
\end{enumerate}


\section{Results}
\label{sec:results}

 \subsection{14P/Wolf}
 \label{sec:res_14P}
 
The rotational lightcurve of comet 14P/Wolf was previously observed in 2004 by \cite{Snodgrass2005}. They determined a rotation rate $P$ = 7.53 $\pm$ 0.10 hours. In \citetalias{Kokotanekova2017}, we revised this period by adding a dataset from 2007, in the same aphelion arc, and derived a rotation period $P$ = 9.02 $\pm$ 0.01 hours. 

We observed 14P again in 2016 in order to look for changes in its spin rate during the last apparition. The new observations in July 2016 were taken almost a full orbit later, while the comet was inbound, after it had passed through perihelion in 2009 and aphelion in 2013. 

\begin{figure}
    \centering
   \includegraphics[width=0.48\textwidth]{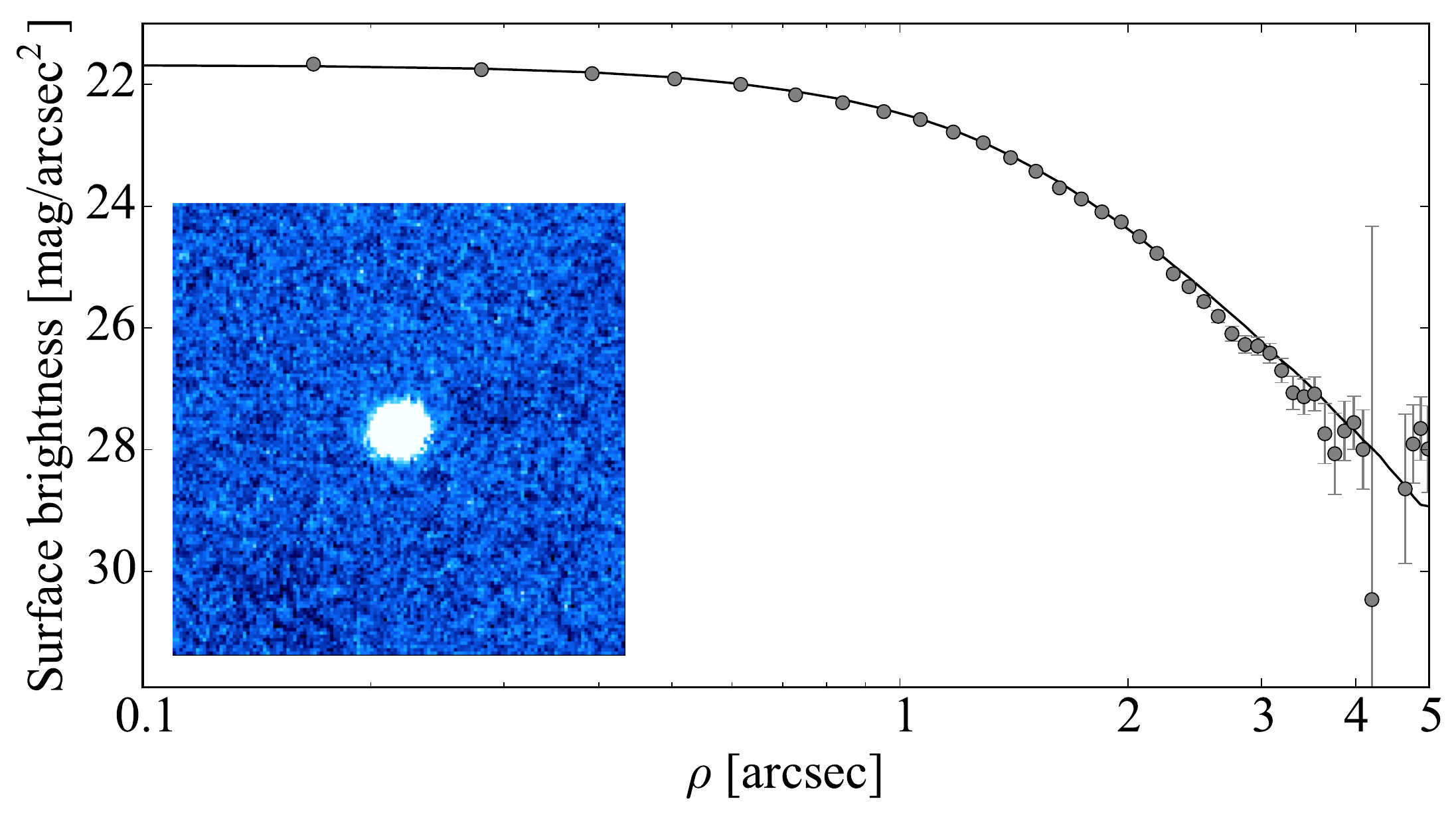}
   \caption{\rknew{Surface brightness profile of comet 14P from 7 July 2016. The image in the lower left shows a 30 $\times$ 30 arcseconds composite image of 14P made of 12 $\times$ 300 s exposures. The frames were added using the method described in \citetalias{Kokotanekova2017}. The comet had a stellar-like profile and no apparent signatures of activity. The surface brightness of the comet is plotted against radius $\rho$ from the comet centre. The agreement of the comet profile with the scaled stellar PSF (solid line), indicates that the comet was observed as a point source, and appeared as inactive during the observations.}}
    \label{14P_2016_PSF}%
    \end{figure}

Comet 14P was observed during five consecutive nights in July 2016 using LAICA on the CAHA 3.5m telescope. \rknew{The comet was inactive during the observations as shown by its stellar profile in the combined image (Fig.~\ref{14P_2016_PSF})}. The phase angle changed by less than 0.6 degrees during the observing run, and therefore the adopted phase function correction is expected to have a negligible effect on the derived rotational lightcurve. 

\begin{figure}
    \centering
   \includegraphics[width=0.48\textwidth]{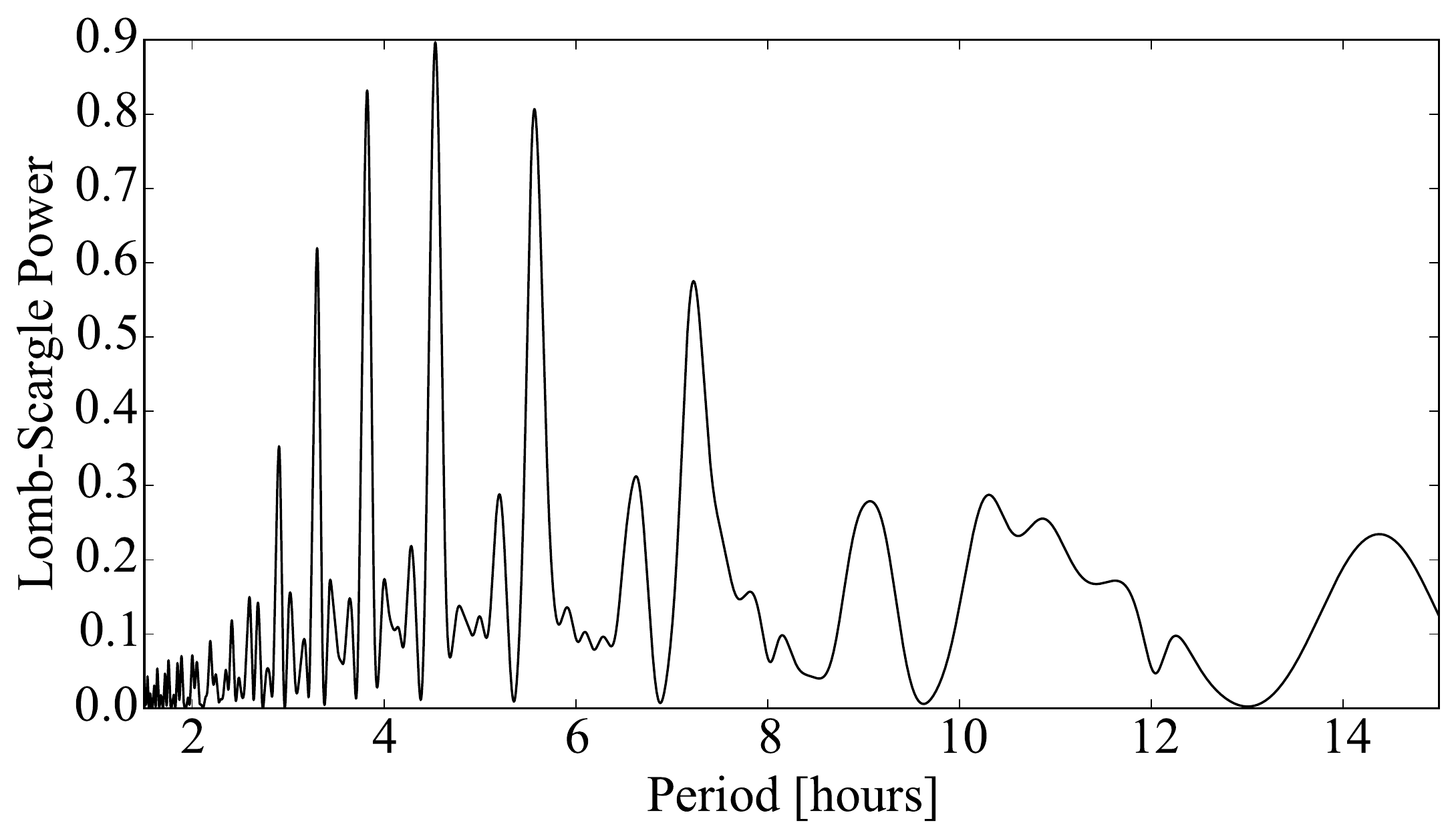}
   \caption{LS periodogram for 14P from the dataset collected in July 2016. The plot shows the LS power versus period. The highest peak occurs at 4.54 which corresponds to a double-peaked lightcurve with period $P$ = 9.07 h.}
    \label{14P_2016_LS}%
    \end{figure}

\begin{figure}
    \centering
   \includegraphics[width=0.48\textwidth]{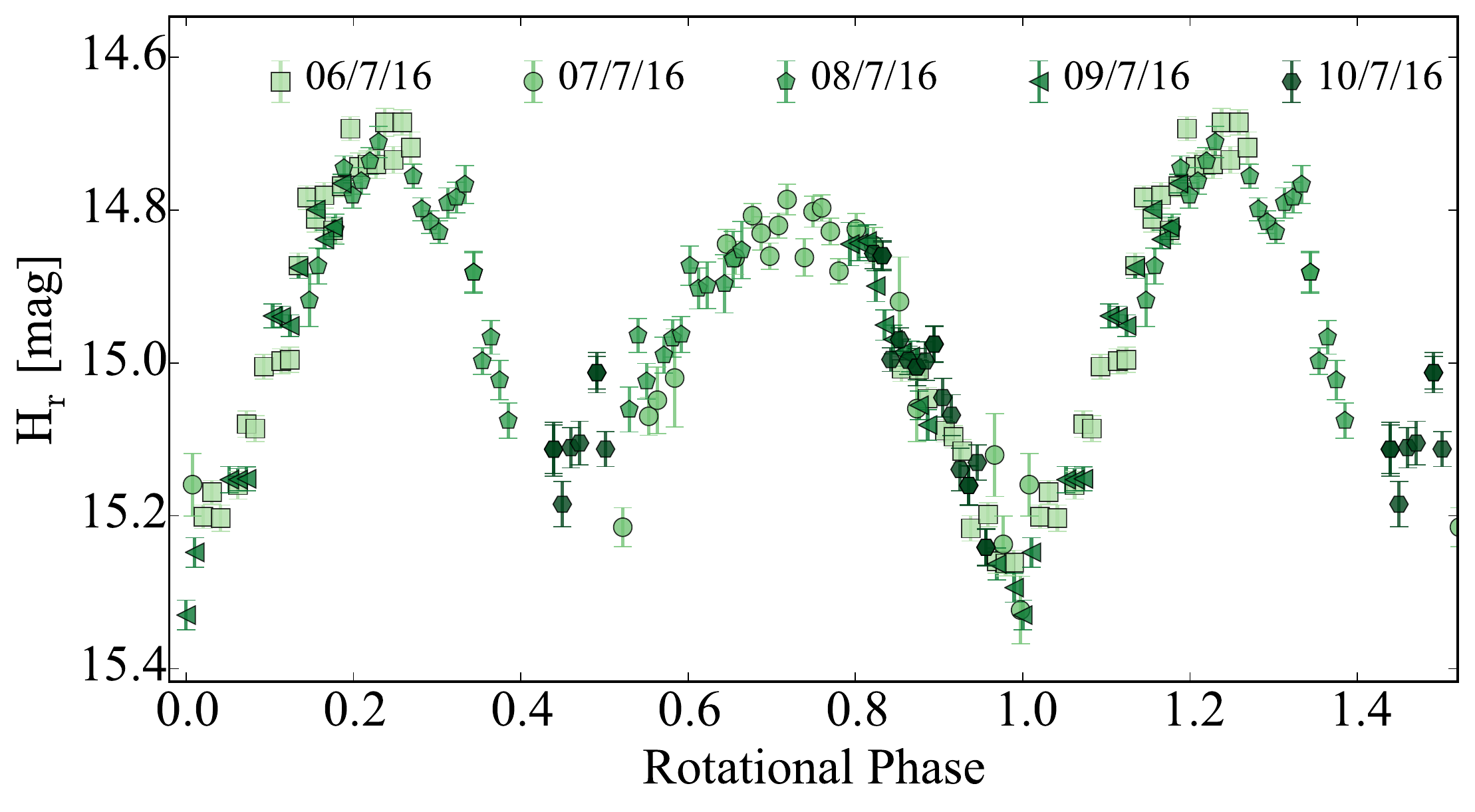}
   \caption{Rotational lightcurve of 14P with the data from 2016. The lightcurve is folded with the LS best period of 9.07 h hours. The error bars indicate the combined 1-$\sigma$ uncertainty of the differential photometry and the absolute photometric calibration.}
    \label{14P_2016_LC}%
    \end{figure}

In \citetalias{Kokotanekova2017}, we found a phase-function slope $\beta$ = 0.060 $\pm$ 0.005 mag/deg for 14P. We used this slope to correct the data, and looked for possible periods. Figure \ref{14P_2016_LS} displays the LS periodogram with a highest peak corresponding to a double-peaked lightcurve with period 9.07 hours. We inspected the lightcurves corresponding to the other two prominent peaks in the LS periodogram, at 7.6 hours and 11.1 hours, but they produced lightcurves with a significantly larger scatter. The lightcurve of 14P phased with the period $P$ = 9.07 hours is plotted in Fig. \ref{14P_2016_LC}. There are data points covering all phases of the lightcurve, and they clearly show that the lightcurve of 14P has asymmetric peaks.  

\begin{figure}
    \centering
   \includegraphics[width=0.48\textwidth]{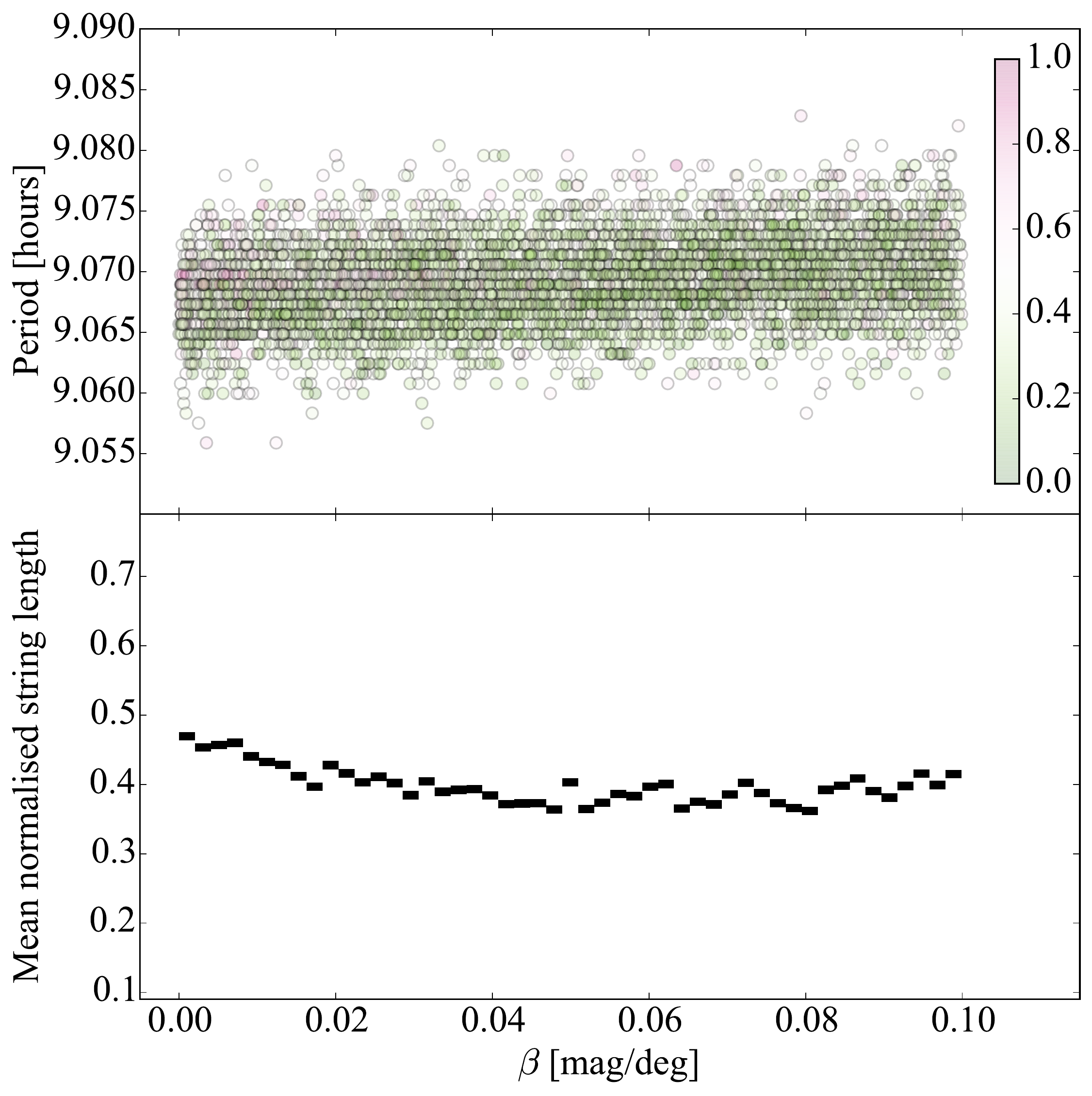}
   \caption{Results from the MC2 method used to determine the range of possible rotation periods of 14P using the 2016 data. The MC2 method looked for periods between 3 and 30 hours using phase-function slopes in the range 0.0 - 0.1 mag/deg. The top panel contains the distribution of the rotation periods derived for each clone. The colour of the points corresponds to the normalised range of the total lightcurve string length computed for each clone. The bottom panel shows the mean of the normalised string length for $\beta$ bins of 0.001 mag/deg width.}
    \label{14P_2016_MC2}%
    \end{figure}

To test the robustness of this period determination, we used the MC2 method to search for rotation periods between 3 and 30 hours. For phase-function slopes in the range from 0.0 to 0.1 mag/deg, we determined that the range of possible solutions is 9.056 - 9.083 hours. The top panel of Fig. \ref{14P_2016_MC2} shows the distribution of all clones from the MC2 run. The derived period range appears to be largely independent of the chosen slope, although a slight trend for longer periods with increasing $\beta$ can be noticed. The colour scale in the plot indicates the goodness of the lightcurve for each clone and corresponds to the normalised string length. The bottom panel of Fig. \ref{14P_2016_MC2} shows that the mean of the string length does not vary significantly. This confirms that we cannot unambiguously determine the phase-function slope from this data set, given the limited range in $\alpha$ of the observations in 2016.  For $\beta$ = 0.060 $\pm$ 0.005 mag/deg derived in \citetalias{Kokotanekova2017}, the range of possible periods is 9.060 - 9.079 hours. We therefore conclude that in July 2016 the rotation rate of 14P was in the range 9.06 - 9.08 hours. 

\rknew{It is possible to estimate the maximum difference between the sidereal ($P_{\mathrm{sid}}$) and synodic ($P_{\mathrm{syn}}$) rotational periods using the following expression from \cite{Pravec1996}:}
\begin{equation}
|P_{\mathrm{sid}}-P_{\mathrm{syn}}| \leq \omega_{\mathrm{PAB}} P_{\mathrm{syn}}^2, 
\end{equation}
\rknew{where $\omega_{\mathrm{PAB}}$ is the angular velocity of the phase angle bisector  \cite[PAB, for a definition, see][]{Harris1984}. Generally, it can be concluded that for the typically large heliocentric distances necessary for the observations of bare comet nuclei, the PAB changes very slowly. For the duration of the observing run in July 2016, we estimated that the difference between the sidereal and the synodic period of comet 14P was less than 0.0001 hours, which is considerably smaller than the uncertainty of our period determination.  }
  
        \begin{figure}
    \centering
   \includegraphics[width=0.48\textwidth]{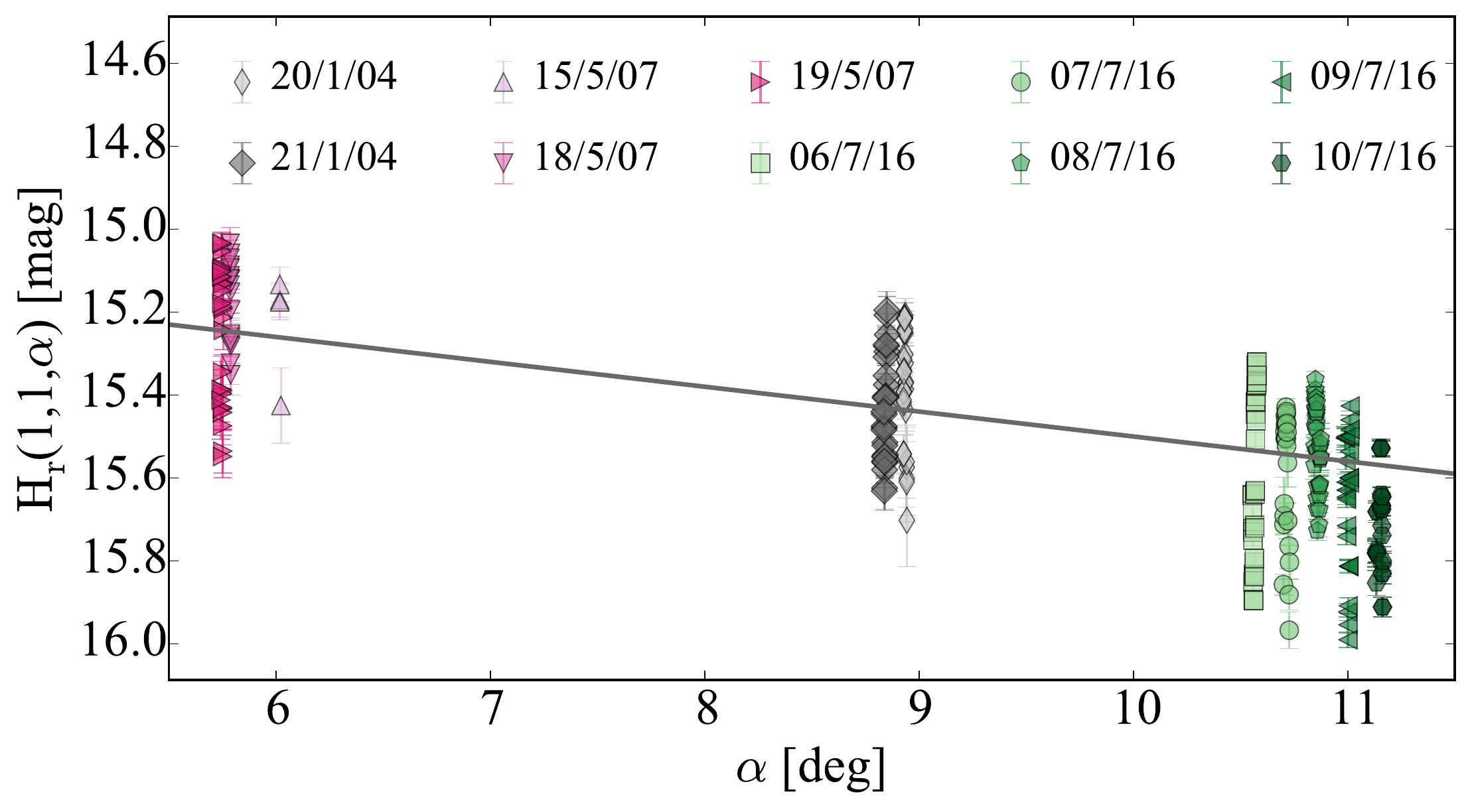}
   \caption{Phase function of comet 14P with the datasets taken in 2004, 2007 and 2016. The calibrated absolute magnitudes of the comet are plotted against phase angle $\alpha$. Over-plotted is a linear phase function model with $\beta$ = 0.060 mag/deg.}
    \label{14P_ALL_PHASE}%
    \end{figure}

    \begin{figure}
    \centering
   \includegraphics[width=0.48\textwidth]{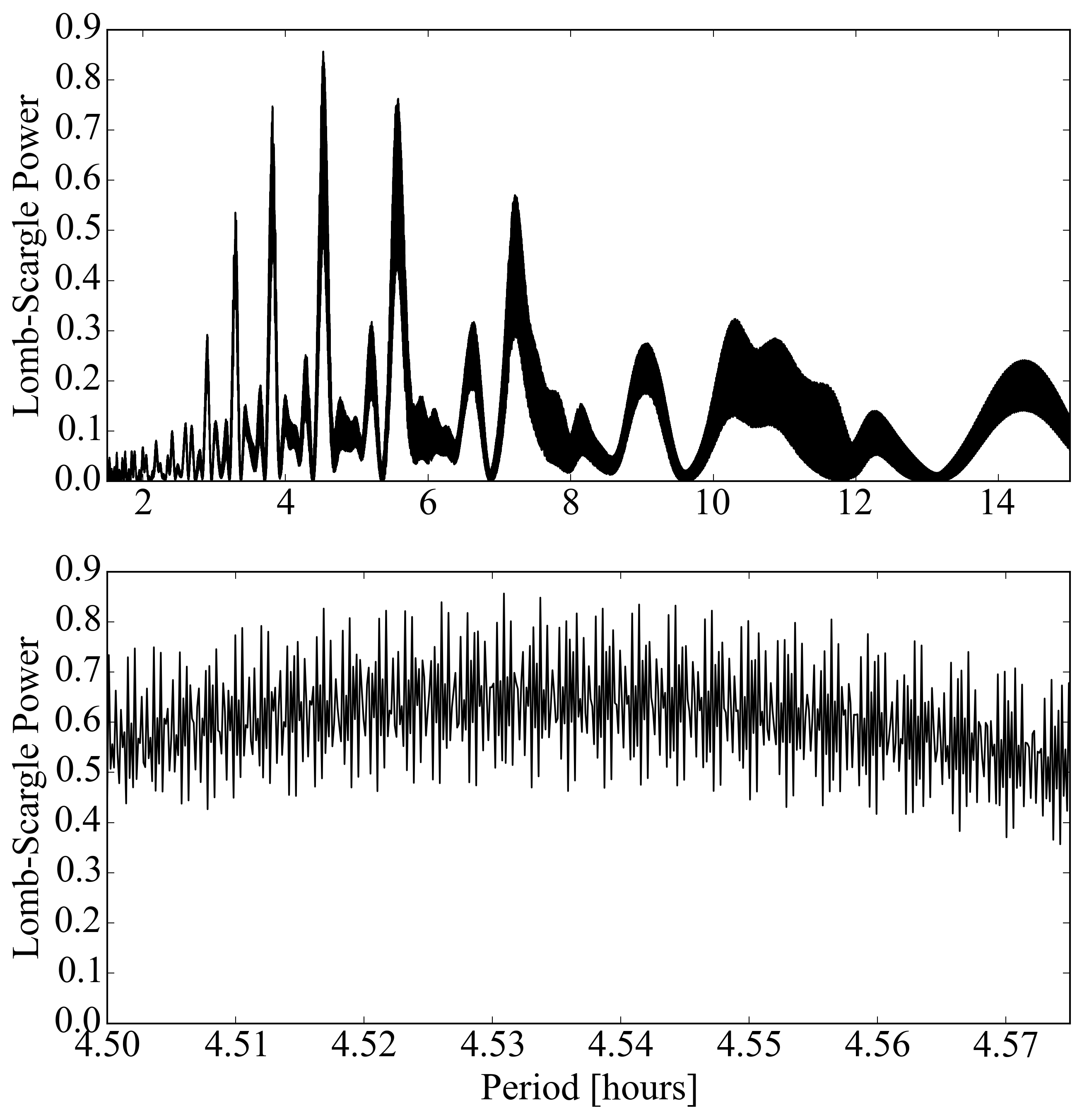}
   \caption{LS periodogram of the combined dataset for 14P collected in 2004, 2007 and 2016 and corrected using a phase-function slope $\beta$ = 0.060 mag/deg. The highest peak corresponds to a period of 9.06748 h, but due to the large timespan between the observing epochs and the resulting aliasing, the periodogram is densely packed with other close-by maxima. The bottom panel shows an enlarged view of the highest peak.}
    \label{14P_ALL_LS}%
    \end{figure}
    
    \begin{figure}
    \centering
   \includegraphics[width=0.48\textwidth]{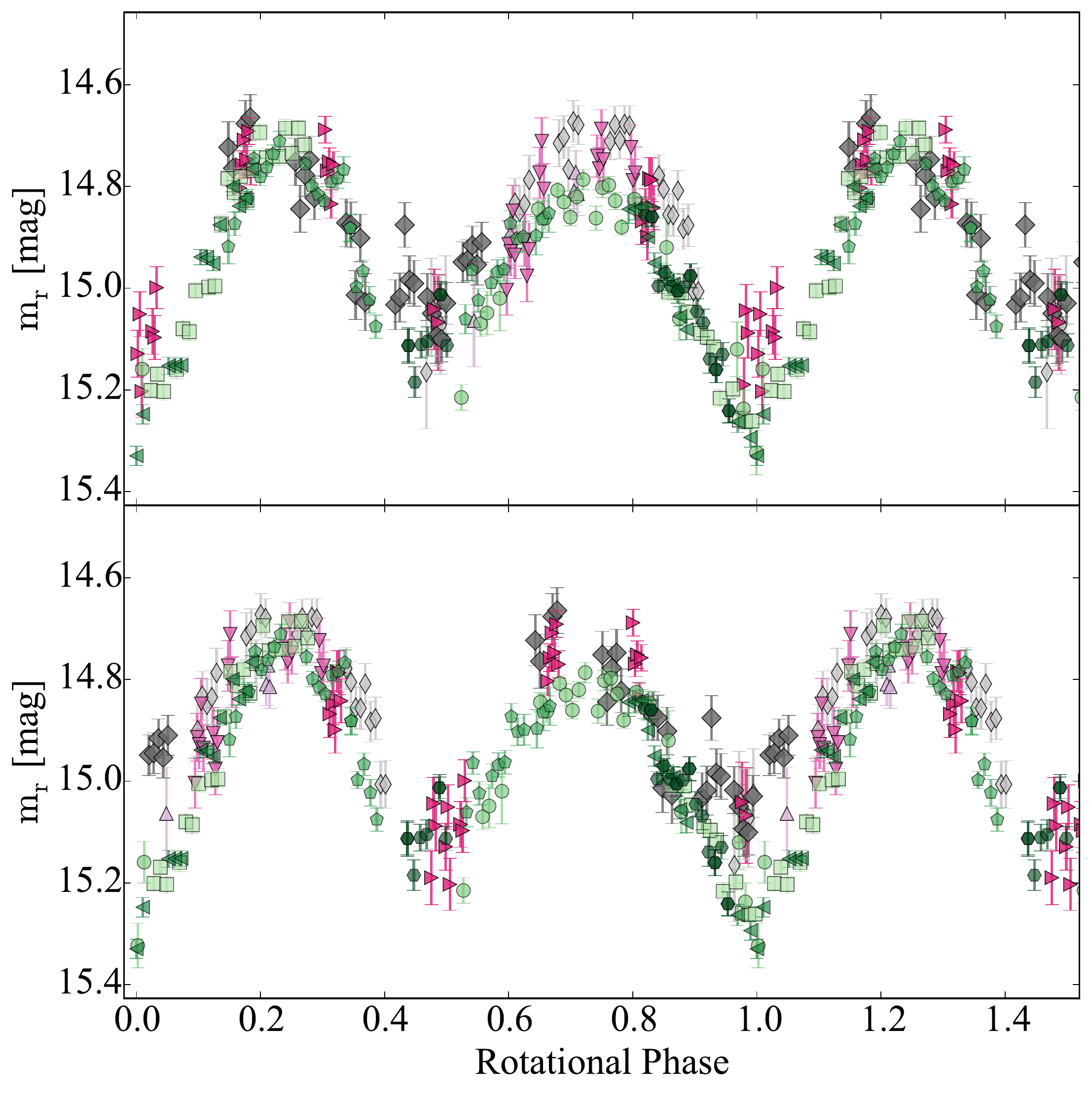}
   \caption{Rotational lightcurve of comet 14P with the combined dataset from 2004, 2007, 2016. The symbols correspond to these used in Fig. \ref{14P_ALL_PHASE}. The data were corrected for a phase-function slope $\beta$ = 0.06 mag/deg and the lightcurves are phased with two of the possible periods according to the LS periodogram: $P_\mathrm{1}$ = 9.07313~h (top) and $P_\mathrm{2}$ = 9.07878 h (bottom). The good alignment of the points from the two apparitions indicates that we can find possible rotation periods which satisfy the observations from all three epochs. In both example lightcurves the points from 2004 deviate from the 2016 data. We interpret the difference in the peak-to-peak amplitudes as a result of changes in the viewing geometry between the two epochs.}
    \label{14P_ALL_LC}%
    \end{figure}
    
    \begin{figure}
    \centering
   \includegraphics[width=0.48\textwidth]{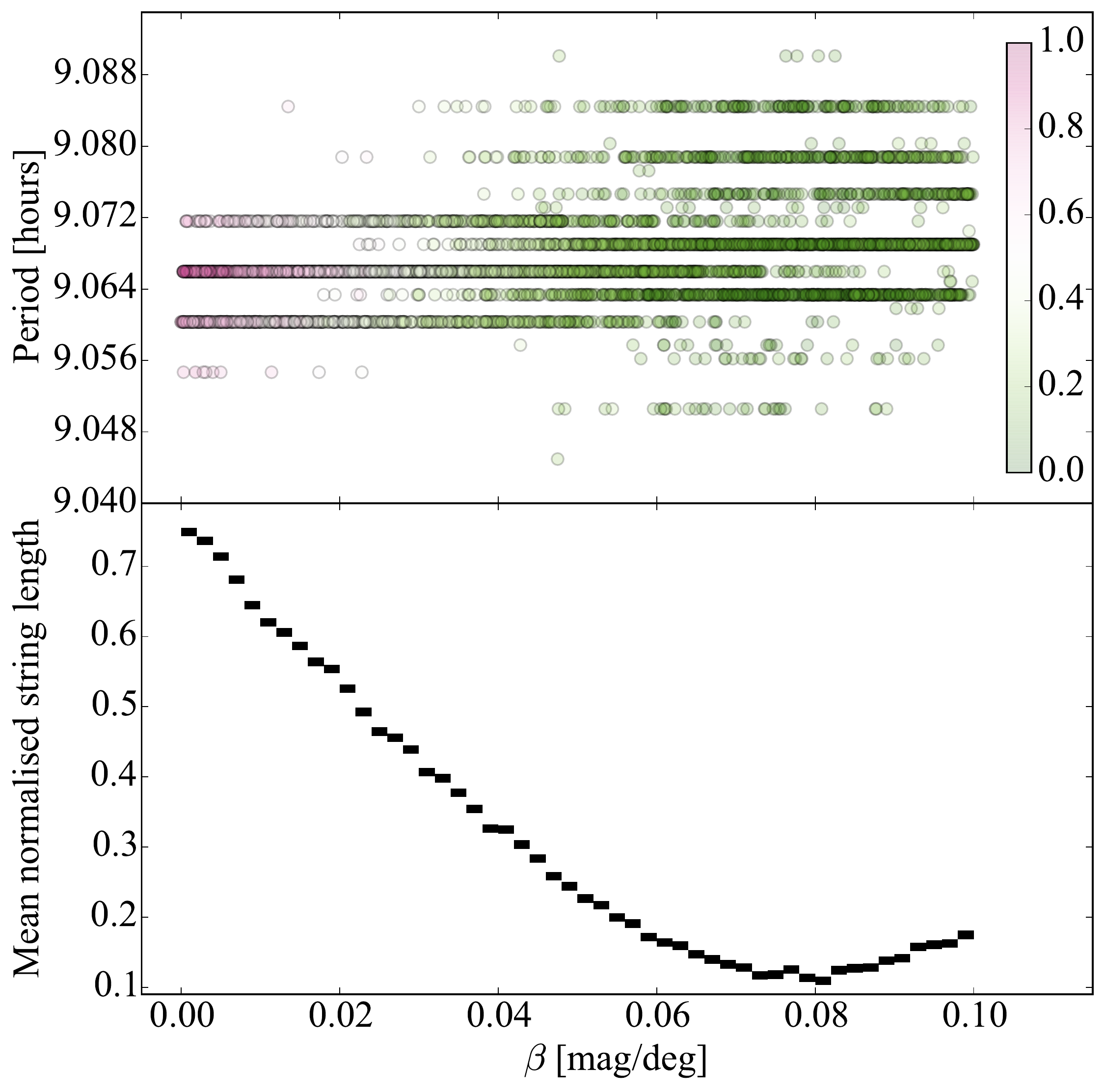}
   \caption{Same as Fig. \ref{14P_2016_MC2} for the 14P data from the combined datasets taken in 2004, 2007 and 2016 data. We assigned a range of possible phase-function slopes of 0.00 - 0.10 mag/deg and looked for periods in the range 8 - 10 hours. This diagram shows that we are able to find common periods for all datasets in the range 9.04 - 9.09 hours. The MC2 method indicates a preference for lightcurves with phase-function slopes between 0.07 and 0.08 mag/deg. }
    \label{14P_ALL_MC}%
    \end{figure}


The lightcurve period derived from the current data set is very close to the period $P$ = 9.02 $\pm$ 0.01 hours from \citetalias{Kokotanekova2017}. If the difference between the two period determinations is taken directly, then it would imply a period change of between 1.8 and 4.2 minutes per orbit. However, before this conclusion is made, it is important to point out that the uncertainty of the two periods was derived from the MC method in \citetalias{Kokotanekova2017} and the MC2 method in this work. While these procedures aim to quantify the uncertainty of the derived periods by taking into account the photometric and calibration uncertainties as well as the phase-function correction, they might not account for all possible solutions. Each of the iterations in the Monte Carlo methods determines only the most likely period from the LS periodogram, and does not consider other less likely but possible periods. This means that the two datasets need to be examined together in order to confirm the period change. 

We therefore attempted to find a common period which would satisfy the data from all three epochs. We looked for possible common rotation periods by combining the old datasets from 2004 and 2007 with the new data from 2016. To correct the data, we used the slope $\beta$ = 0.060 mag/deg (Fig. \ref{14P_ALL_PHASE}). The resulting LS periodogram in Fig. \ref{14P_ALL_LS} has a maximum at around 9.07 hours, but a careful inspection shows the presence of many aliases due to the large timespan between the observations.

In Fig. \ref{14P_ALL_LC}, we have plotted lightcurves with two of the many possible periods suggested by the LS periodogram. These lightcurves show that it is possible to find common periods for the lightcurves from the two epochs. We can therefore conclude that, given the current set of observations, we cannot detect period changes between the two apparitions. The currently available data do not allow us to rule out changes, and we therefore give the maximum change derived above as an upper limit, i.e. $\Delta P < 4.2$ minutes, but the default conclusion given the existence of a common period to all data should be that the period did not change.

However, it is important to note that the match between the separate lightcurves is not perfect. There are differences in the maximum peaks and the depth of the minima between the data from 2004 and 2016 (Fig. \ref{14P_ALL_LC}). We interpret these differences as a result of change in the viewing geometry -- a different observer latitude, based on the relative orientation of the comet rotation pole and the line of sight to Earth, implying a different lightcurve amplitude -- rather than as evidence for a period change. 

We applied the MC2 procedure to the combined data set for a phase function range of 0.0 - 0.1 mag/deg, and looked for periods in the range 8 - 10 hours. The distribution of possible periods from Fig. \ref{14P_ALL_MC} indicates that the total range of possible common periods for the combined data set from the two apparitions is 9.04 - 9.09 hours. 

According to the results from the MC2 method in Fig. \ref{14P_ALL_MC}, the periods with shortest string lengths are found around 9.062 hours and with phase-function slopes between 0.07 and 0.08 mag/deg. This would imply that the phase-function slope of 14P is steeper than the previously determined value of $\beta$ = 0.060 $\pm$ 0.005 from \citetalias{Kokotanekova2017}. Looking at Fig.~\ref{14P_ALL_PHASE}, it can be seen that the 2016 data are taken at larger phase angle and are, on average, below the previously identified trend, which explains the steeper slope found when including these data. The best slope from the MC2 method is derived under the assumptions that the spin rate of the comet has remained constant and that the different viewing geometry does not have a large effect on the observed lightcurve. Since both of these assumptions might be false, we consider the value of $\beta$ = 0.060 $\pm$ 0.005 mag/deg to be a better estimate of the phase-function slope since it was derived from observations taken during the same orbit around the Sun.


 \subsection{143P/Kowal-Mrkos}
 \label{sec:res_143P}
  
The rotation rate of comet 143P was first determined from observations in 2001 by \cite{Jewitt2003}. They derived a period $P$ = 17.21 $\pm$ 0.10 hours and a phase-function slope $\beta$ = 0.043 $\pm$ 0.014 mag/deg. Since then the comet has passed perihelion once, in June 2009, which motivated us to search for possible spin-rate changes that may have resulted from the comet's activity.

\begin{figure}
    \centering
   \includegraphics[width=0.48\textwidth]{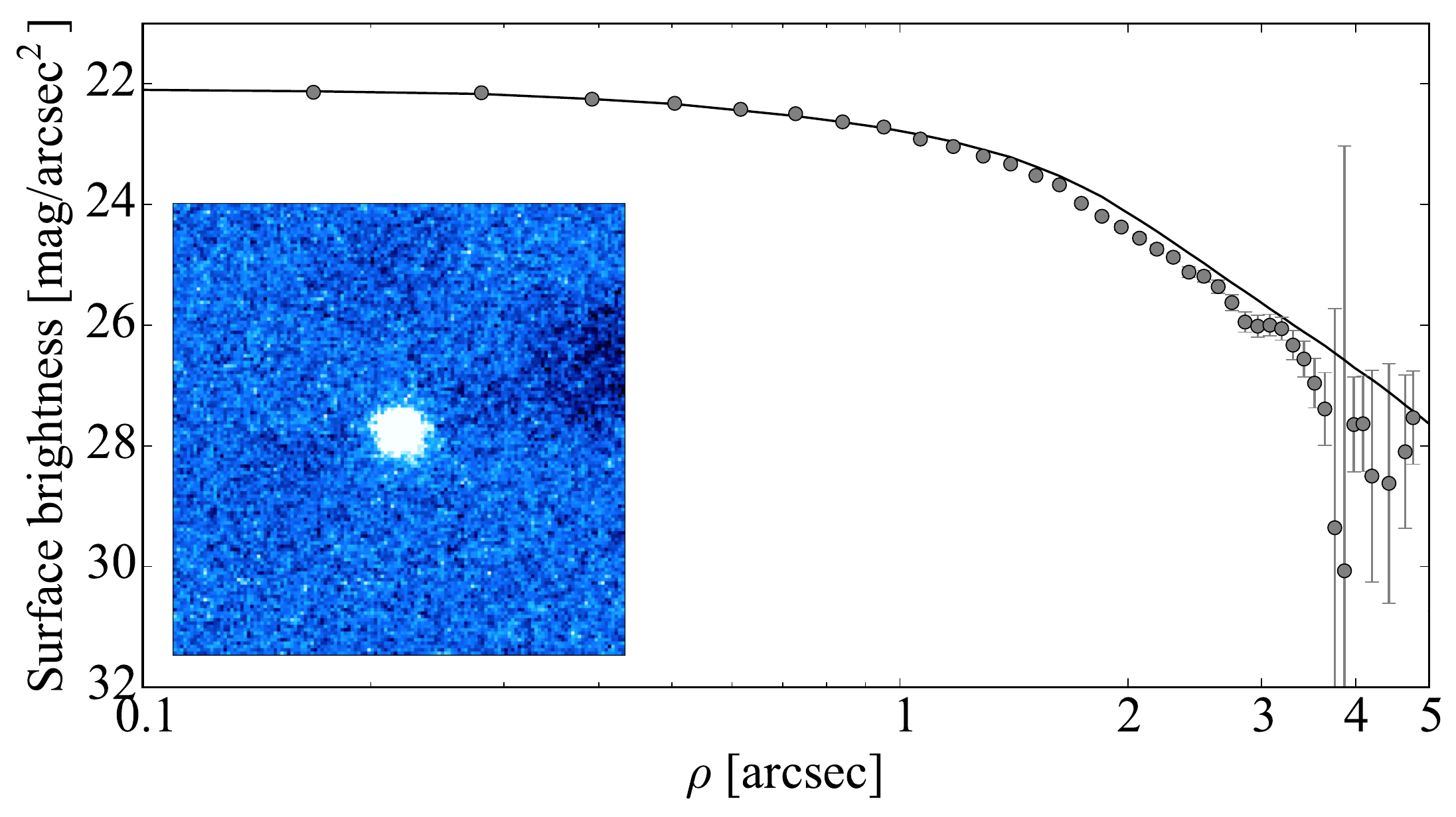}
   \caption{\rknew{Same as Fig. \ref{14P_2016_PSF}, for the observations of 143P from 16 January 2016. The composite image in the lower left corner was made up of 15 $\times$ 180 s exposures. The stellar appearance in the composite image and the agreement of the surface brightness profile of the comet with the stellar PSF suggest that the comet was inactive during the observations in 2016.}}
    \label{143P_2016_PSF}%
    \end{figure}

\begin{figure}
    \centering
   \includegraphics[width=0.48\textwidth]{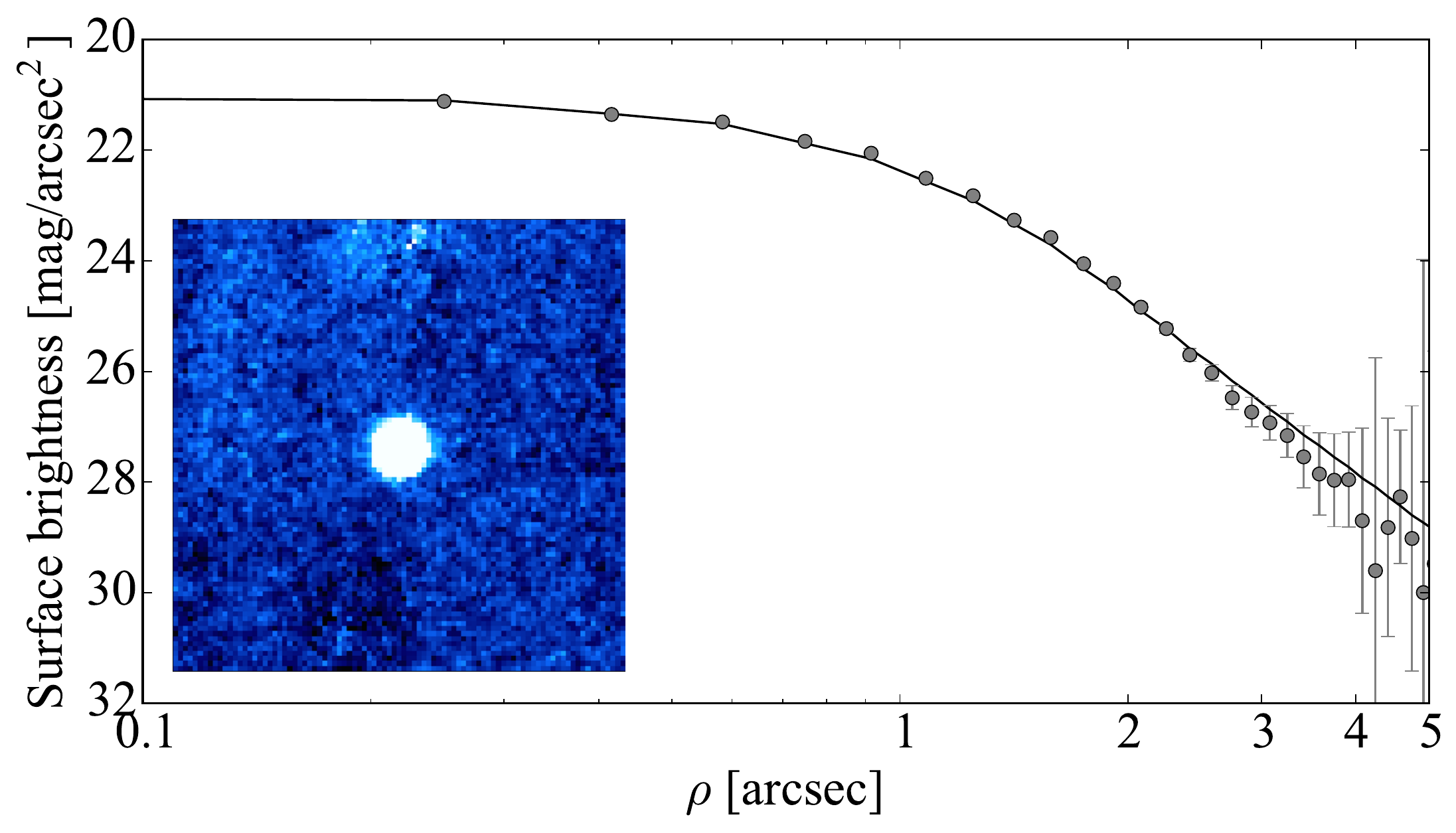}
   \caption{\rknew{Same as Fig. \ref{14P_2016_PSF}, for the observations of 143P from 18 February 2017. The composite image in the lower left corner was made up of 14 $\times$ 180 s exposures. 
   }}
    \label{143P_2017_PSF}%
    \end{figure}

We made two attempts to observe the rotational lightcurve of 143P while it was inbound. In January 2016 we observed 143P with LAICA on the 3.5-meter telescope at Calar Alto. In February and March 2017 we used INT and the Rozhen 2-meter telescope. \rknew{The comet did not show signs of activity during the observations (Figs. \ref{143P_2016_PSF} and \ref{143P_2017_PSF})}. Therefore, due to the lack of outgassing, its rotation rate most likely remained unchanged between 2016 and 2017, and we proceeded to combine the two epochs in order to derive the current rotation rate of 143P. 

\begin{figure}
    \centering   
    \includegraphics[width=0.48\textwidth]{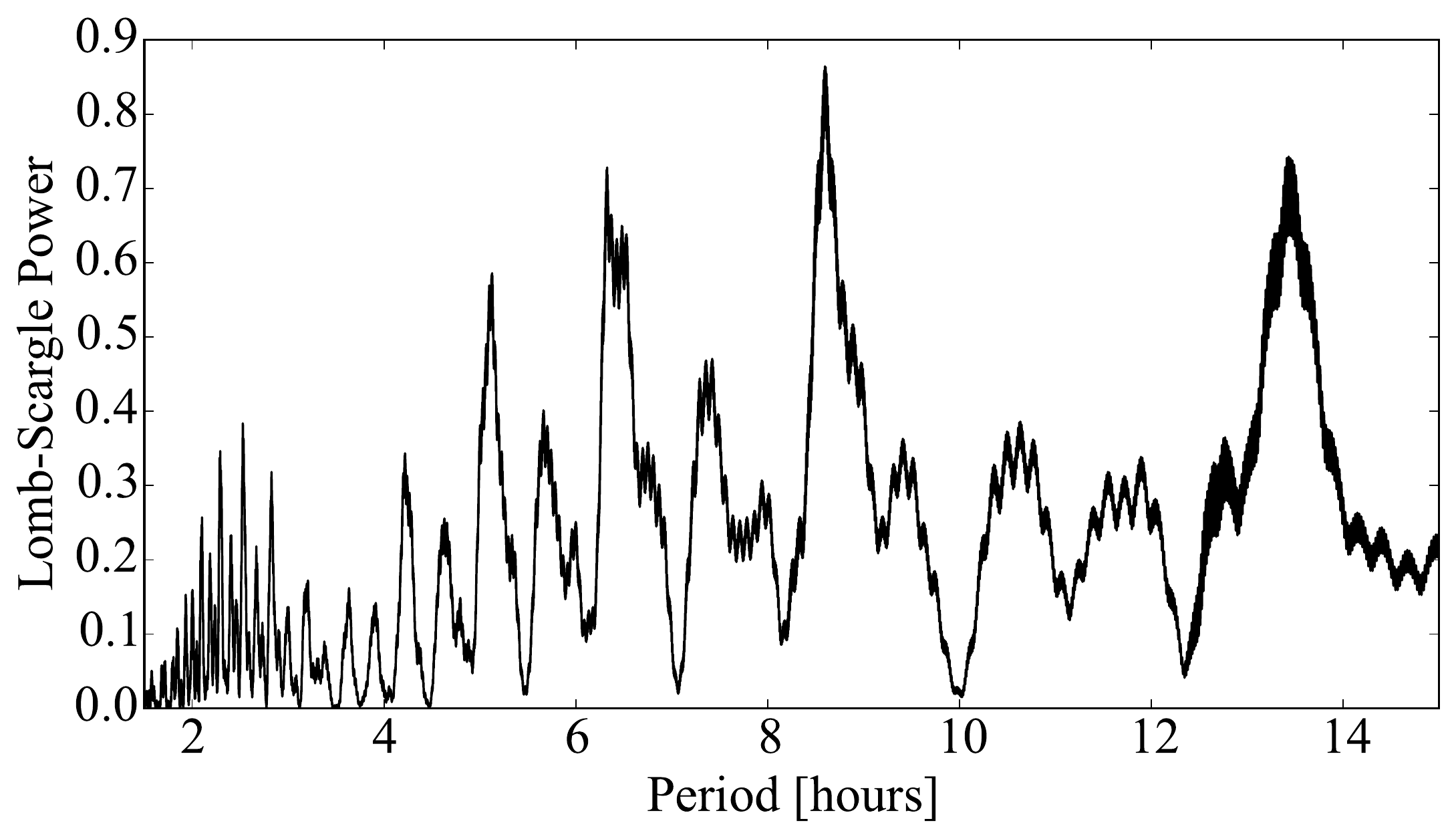}
   \caption{LS periodogram for 143P from the dataset collected in 2016 and 2017, and corrected for a phase-function slope $\beta$ = 0.043 mag/deg. The plot shows the LS power versus period. The highest peak corresponds to a double-peaked lightcurve with period $P$ = 17.197 h.}
    \label{143P_NEW_LS}%
    \end{figure}

As a first step we corrected the new data with the phase-function slope $\beta$ = 0.043 $\pm$ 0.014 mag/deg from \cite{Jewitt2003}. We then inspected the LS periodogram of the combined dataset (Fig. \ref{143P_NEW_LS}). The periodogram indicated a maximum corresponding to a period of $\sim$ 17.197 hours but suffered from aliasing due to the time gaps in the observations.

\begin{figure}
    \centering
   \includegraphics[width=0.48\textwidth]{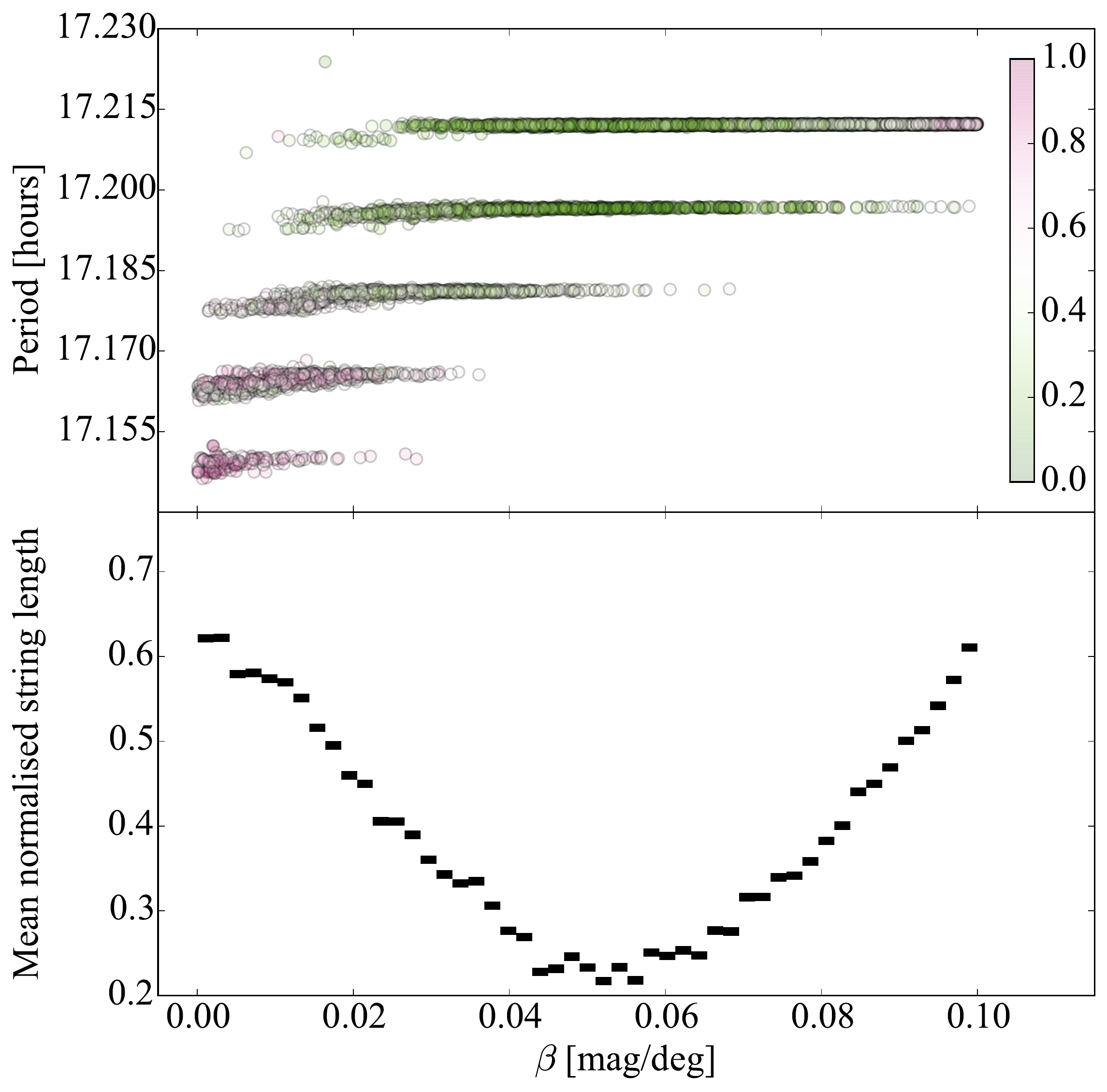}
   \caption{Results from the MC2 method applied to the 143P data from the combined datasets taken in 2016 and 2017. The MC2 method was run for a range of possible phase-function slopes $\beta$ = 0.00 - 0.10 mag/deg and periods from 3 to 30 hours.  }
    \label{143P_NEW_MC}%
    \end{figure}

\begin{figure}
    \centering
  \includegraphics[width=0.48\textwidth]{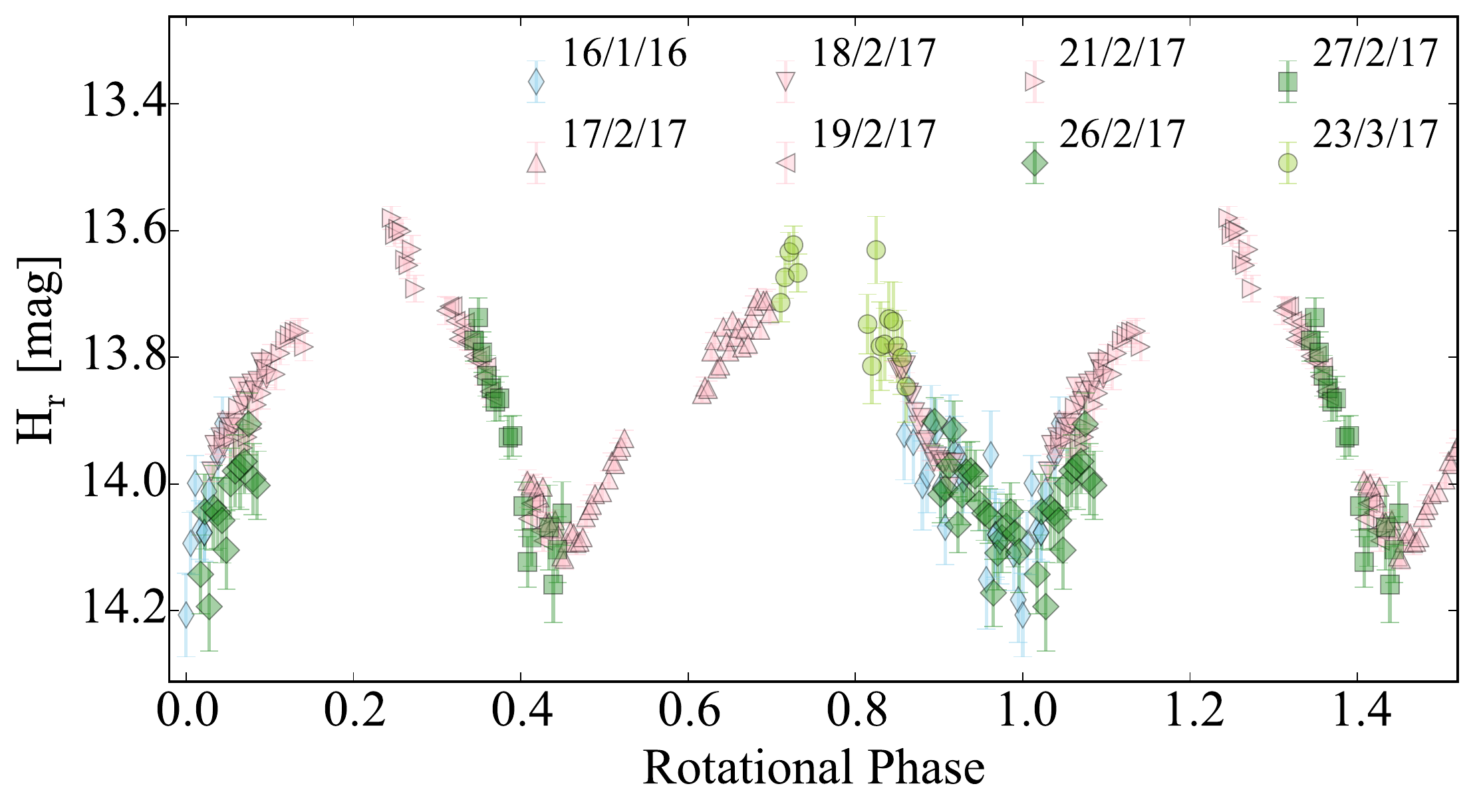}
   \caption{Rotational lightcurve of comet 143P from the data taken in 2016 and 2017. The magnitudes from 17-21 February and from 26-27 February were derived using the same same set of comparison stars and are therefore plotted in the same colours. This lightcurve was corrected for a phase-function slope $\beta$ = 0.051 mag/deg and was phased with a period $P$ = 17.1966 hours, and corresponds to the best lightcurve from the MC2 test. }
    \label{143P_LC_NEW}%
    \end{figure}

In order to derive a common period for the data from 2016 and 2017, we used the MC2 method for phase-function slopes in the range 0.0-0.1 mag/deg and searched for periods between 3 and 30 hours. The results of the MC2 test can be seen in Fig. \ref{143P_NEW_MC}. The possible solutions for the full phase-function slope range between 17.145 and 17.22 hours. As the lower panel in Fig. \ref{143P_NEW_MC} shows, the best lightcurves are found around slope $\beta$ = 0.05 mag/deg. A careful inspection of the results suggests that the clones with phase-function slopes $\beta$ $<$ 0.3 mag/deg, $\beta$ $>$ 0.7 mag/deg and $P$ $<$ 17.18 hours produce lightcurves with a large scatter. Therefore, we conclude that the rotation rate of comet 143P is between 17.18 and 17.22 hours, at one of the following distinct periods: 17.1966 $\pm$ 0.0003, 17.2121 $\pm$ 0.0002 and 17.1812 $\pm$ 0.0002. In Fig. \ref{143P_LC_NEW} we have plotted the best lightcurve according to the MC2 test. The observations cover the whole lightcurve phase and provide very good coverage of both minima.

The possible period range of 17.18 - 17.22 hours which we constrained for the current apparition also includes the period $P$ = 17.21 $\pm$ $0.10$ hours from the 2001 data \citep{Jewitt2003}. This implies that no period change was detected between the two epochs, with an upper limit of 6.6 minutes per orbit, largely due to the uncertainty quoted on the 2001 period. 

\begin{figure}
    \centering
  \includegraphics[width=0.48\textwidth]{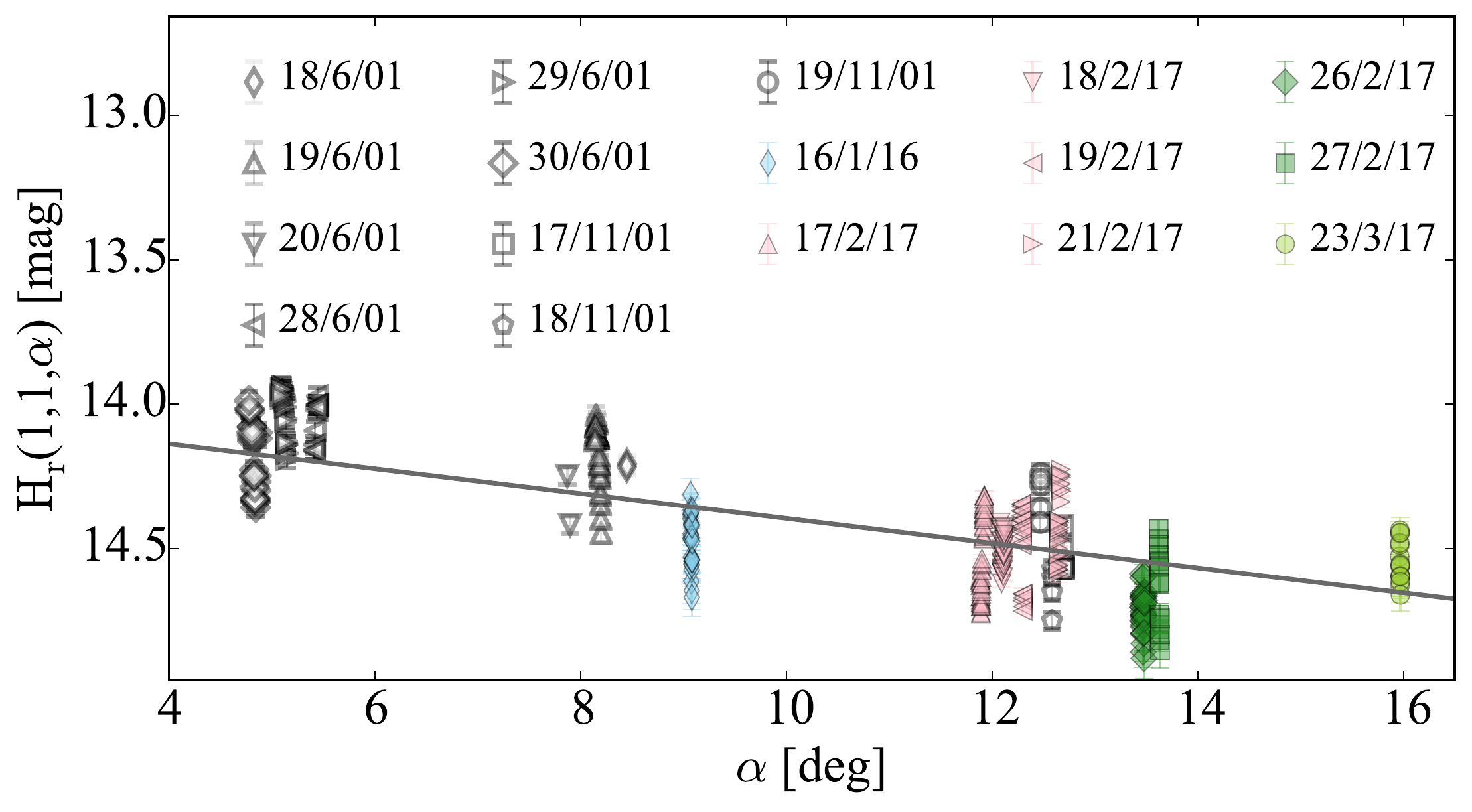}
   \caption{Phase function of comet 143P from the datasets taken in 2001 \protect\citep{Jewitt2003}, 2016 and 2017. The calibrated absolute magnitudes of the comet are plotted against phase angle $\alpha$. The points from 17-21 February 2018 and those from 26-27 February are plotted in the same colours since they were calibrated using the same comparison stars. The absolute magnitudes for 2001 are taken from Table 2 in \protect\cite{Jewitt2003}, and were converted to PS1 $\mathrm{r_{P1}}$-band. Over-plotted is a linear phase function with slope $\beta$ = 0.043 mag/deg from \protect\cite{Jewitt2003}.}
    \label{143P_ALL_PHASE}%
    \end{figure}

To test this conclusion, we used the data points from \cite{Jewitt2003} in order to check whether the lightcurves from the two epochs are consistent, as well as to set an upper limit on a possible period change which might have remained undetected. We converted the magnitudes from \cite{Jewitt2003} to the PS1 $\mathrm{r_{P1}}$-band using the nucleus colour B--V = 0.82 $\pm$ 0.02 mag from \cite{Jewitt2003} and the colour conversion terms from \cite{Tonry2012}. All absolute magnitudes are plotted versus phase angle in Fig. \ref{143P_ALL_PHASE}. The data from \cite{Jewitt2003} show a very good agreement with the new points from this work, and the old phase function $\beta$ = 0.043 $\pm$ 0.014 mag/deg aligns well with the extended dataset.  

\begin{figure}
    \centering
 \includegraphics[width=0.48\textwidth]{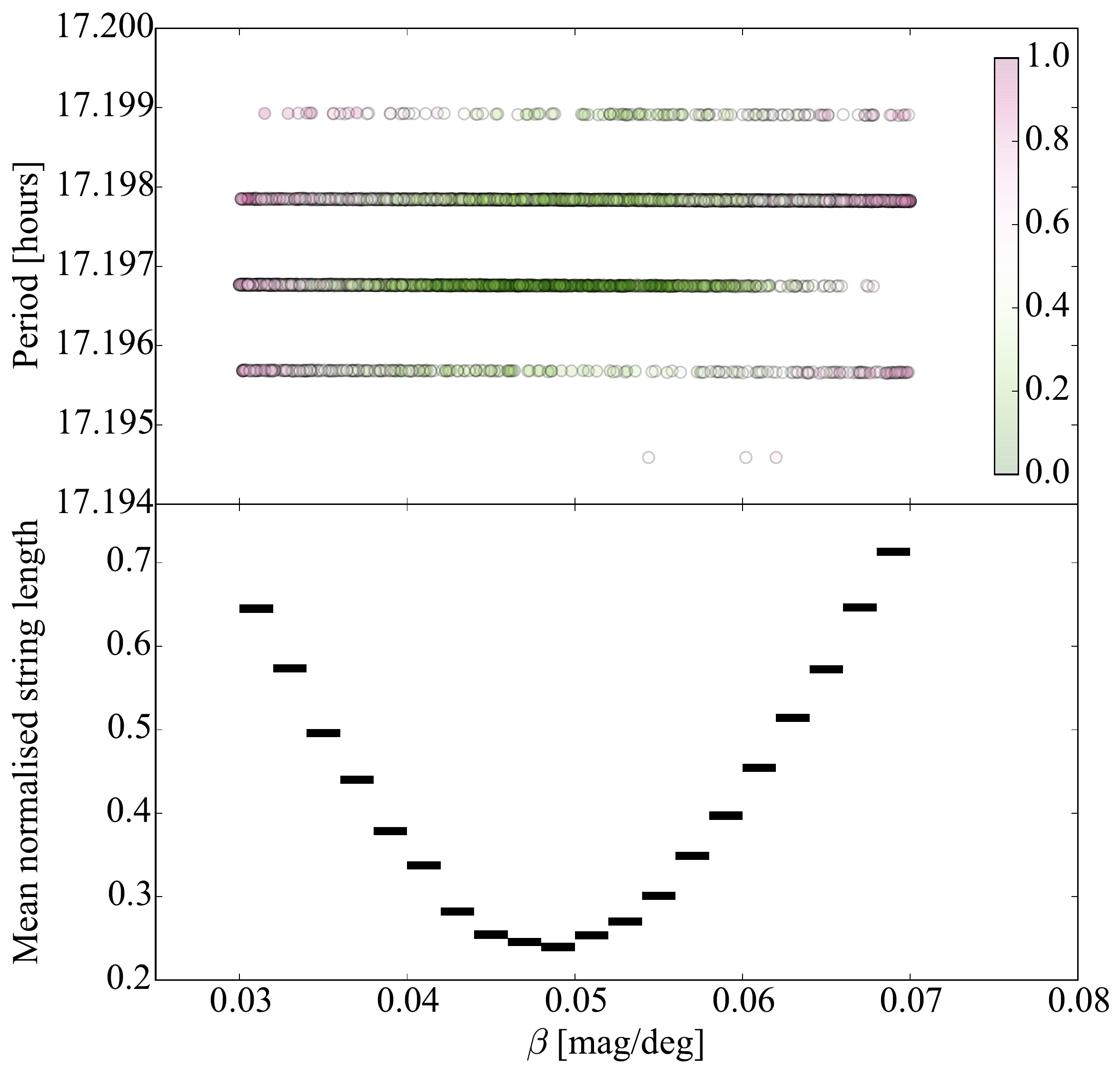}
   \caption{Results from the MC2 method applied to the 143P data from the combined datasets taken in 2001 \protect\citep{Jewitt2003}, 2016 and 2017. The MC2 method was run for a range of possible phase-function slopes $\beta$ = 0.03 - 0.07 mag/deg and periods from 17.18 to 17.22 hours.  }
    \label{143P_ALL_MC}%
    \end{figure}

We next applied the MC2 method to look for common rotation periods of the combined data from 2001, 2016 and 2017. We limited the MC2 test to $\beta$ between 0.03 and 0.07 mag/deg and periods between 17.18 and 17.22 hours, derived for the new dataset above. The MC2 test in Fig. \ref{143P_ALL_MC} identified that the possible common periods lie in the range 17.1945-17.200 hours.

\begin{figure}
    \centering
  \includegraphics[width=0.48\textwidth]{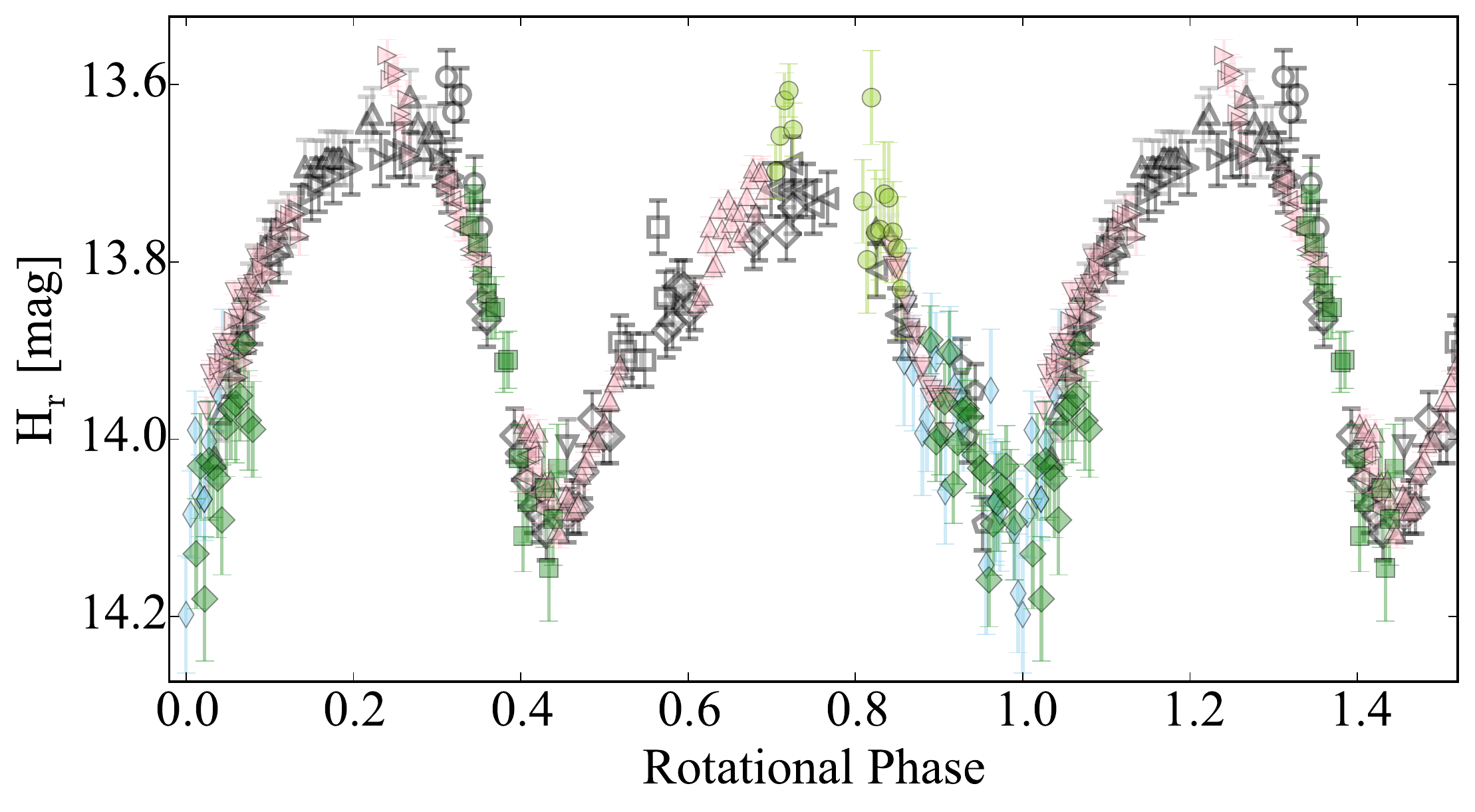}
   \caption{Rotational lightcurve of 143P with the datasets from 2001, 2016 and 2017. The symbols correspond to the ones in Fig. \ref{143P_ALL_PHASE}. The data were corrected with a phase-function slope $\beta$ = 0.052 mag/deg and folded with a period $P$ = 17.19676 h. Those values were selected from the best lightcurves in the output of the MC2 method. }
    \label{143P_LC_ALL}%
    \end{figure}

In Fig. \ref{143P_LC_ALL} we have plotted the common lightcurve with the best phase-function slope and period identified by the MC2 test. This lightcurve illustrates well the remarkable match between the datasets from the two apparitions. While there might be a shift in magnitude between the two datasets due to the different absolute calibration methods used by \cite{Jewitt2003} and here, we were able to derive a well-aligned common lightcurve by varying the phase-function slope. The phase-function slope derived here depends on the assumptions that 1) the absolute calibration from \cite{Jewitt2003} is very precise; 2) changes in the observing geometry (pole position) are negligible; 3) the rotation period of the comet did not change between the two epochs and therefore we are able to derive a common lightcurve. With all of these caveats in mind, we consider the slope $\beta$ = 0.043 $\pm$ 0.014 mag/deg from \cite{Jewitt2003} to be a more reliable estimate, since it uses a broad range of phase angles and was derived from consistently calibrated magnitudes measured during the same apparition.

The radius $R_{\mathrm{n}}$ = $\mathrm{4.79^{+0.32}_{-0.33}}$ km of comet 143P was determined from thermal infrared measurements in 2007 \citep{Fernandez2013}. We use this size together with the absolute magnitude from the lightcurve observations to determine its albedo. 

\cite{Jewitt2003} determined an absolute magnitude $H_{\mathrm{R}}$(1,1,0) = 13.49 $\pm$ 0.20 mag and (B-V) = 0.82 $\pm$ 0.02 mag, which can be converted to $H_{\mathrm{r_{P1}}}$(1,1,0) = 13.70 $\pm$ 0.20 mag. From this magnitude we calculate a geometric albedo $p_{\mathrm{r}}$ = 0.055 $\pm$ 0.013 using:
\begin{equation}
\label{eq:albedo}
p_{\mathrm{r_{P1}}} = (k^2 \, / \ R_{\mathrm{n}}^2 ) \times 10^{0.4(m_{\sun} - H_{\mathrm{r_{P1}}})}.
\end{equation}
In this expression, $m_{\sun} = -26.91$ mag is the apparent magnitude of the Sun in $\mathrm{r_{P1}}$-band and k = 1.496 $\times$ $10^{8}$ km is the conversion factor between au and km. 

This value of the geometric albedo agrees with the conservative albedo estimate which we can derive from our observations from 2016 and 2017. For the broad range of possible $\beta$ from the MC2 test in Fig. \ref{143P_NEW_MC}, 0.03--0.07 mag/deg, we determine an absolute magnitude $H_{\mathrm{r_{P1}}}$(1,1,0) = 13.86 $\pm$ 0.12. For the radius from \cite{Fernandez2013}, this converts to $p_{\mathrm{r_{P1}}}$ = 0.048	$\pm$ 0.009. Since the new dataset was calibrated with our method for precise absolute calibration using the Pan-STARRS catalogue, and is therefore directly comparable to the other comets whose albedos were derived in \citetalias{Kokotanekova2017}, we adopt this value below in Section \ref{sec:discussion}.

It is important to note that the optical observations from 2001, 2016 and 2017 were not taken simultaneously to the infrared data used to determine the size \citep{Fernandez2013}. However, the low activity of 143P \citep[e.g. ][]{Jewitt2003} suggests that it does not undergo significant mass loss and its radius has most likely remained unchanged. Additionally, the very good match between the lightcurves from 2001 and 2016-2017 suggest that the changing viewing geometry does not significantly change the estimated absolute optical magnitude of the comet. Therefore, the derived albedo is considered to be a good estimate.

 \subsection{162P/Siding Spring}
 \label{sec:res_162P}
 
The lightcurve of comet 162P was previously studied from two datasets taken in 2007 and 2012, during two consecutive aphelion passages \citepalias{Kokotanekova2017}. The data from 2012 were collected between April and June 2012 and covered a sufficient phase angle range to allow a phase function determination with $\beta$ = 0.039 $\pm$ 0.02 mag/deg \citepalias{Kokotanekova2017}. The two datasets did not show any evidence for a period change during the perihelion passage between 2007 and 2012, although this could be due to the relatively poor sampling of the lightcurve from 2007. The best period derived for 2012 was 32.852 hours, and for the combined data set, the MC method in \citepalias{Kokotanekova2017} resulted in a common period of 32.853 $\pm$ 0.002 hours. 

\begin{figure}
    \centering
   \includegraphics[width=0.48\textwidth]{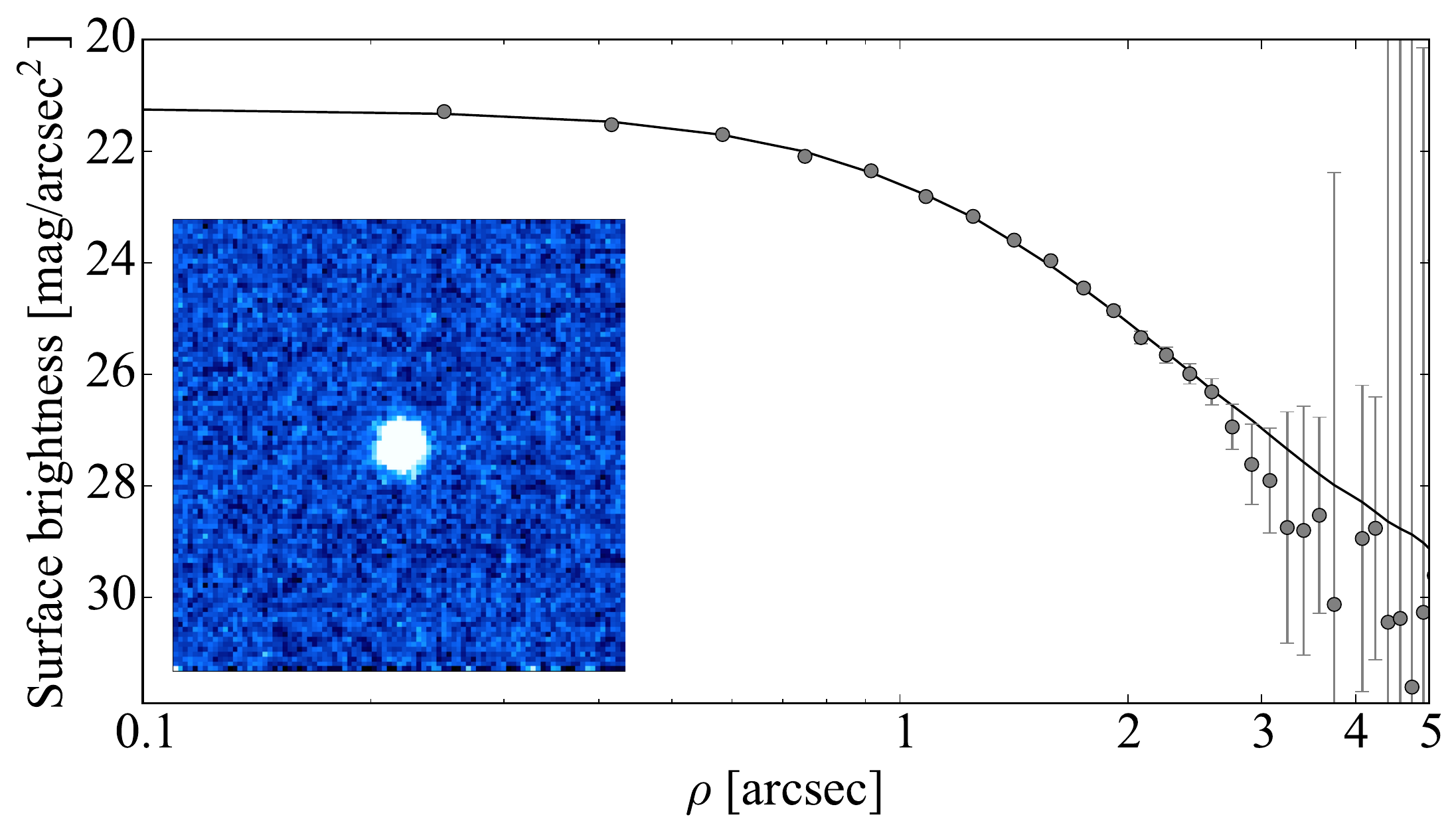}
   \caption{\rknew{Same as Fig. \ref{14P_2016_PSF}, for the observations of 162P from 18 February 2017. The composite image in the lower left corner was made up of 9 $\times$ 120 s exposures. 
   }}
    \label{162P_2017_PSF}%
    \end{figure}

In February 2017 we observed comet 162P during three nights with WFC on INT and one night with FoReRo on the Rozhen 2-meter telescope. These observations were done  before aphelion, almost a full orbit after the previous dataset was taken in 2012. \rknew{Careful analysis of the data from each run determined that the comet was inactive during the observing period (Fig \ref{162P_2017_PSF}).}

The data covered a phase-angle range of approximately 2 degrees, which was insufficient for an independent derivation of the phase function. Therefore, we used the slope $\beta$ = 0.039 $\pm$ 0.02 mag/deg from \citetalias{Kokotanekova2017} to correct the data. 

\begin{figure}
    \centering   \includegraphics[width=0.48\textwidth]{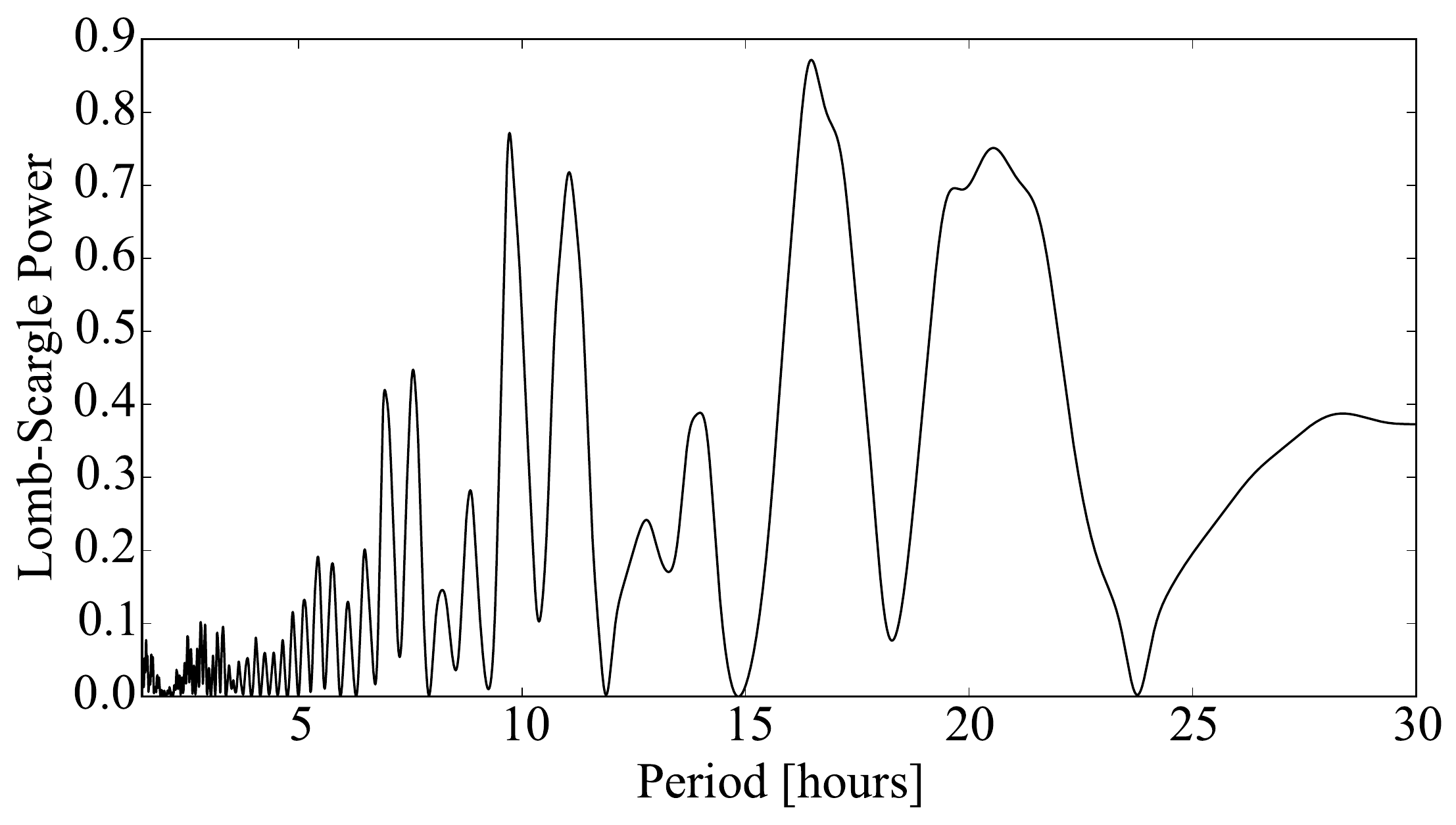}
   \caption{LS periodogram for 162P from the dataset collected in 2017 and corrected with a phase-function slope $\beta$ = 0.039 $\pm$ 0.02 mag/deg. The plot shows the LS power versus period. The highest peak corresponds to a double-peaked lightcurve with period $P$ = 32.92 h.}
    \label{162P_NEW_LS}%
    \end{figure}

\begin{figure}
   \centering
  \includegraphics[width=0.48\textwidth]{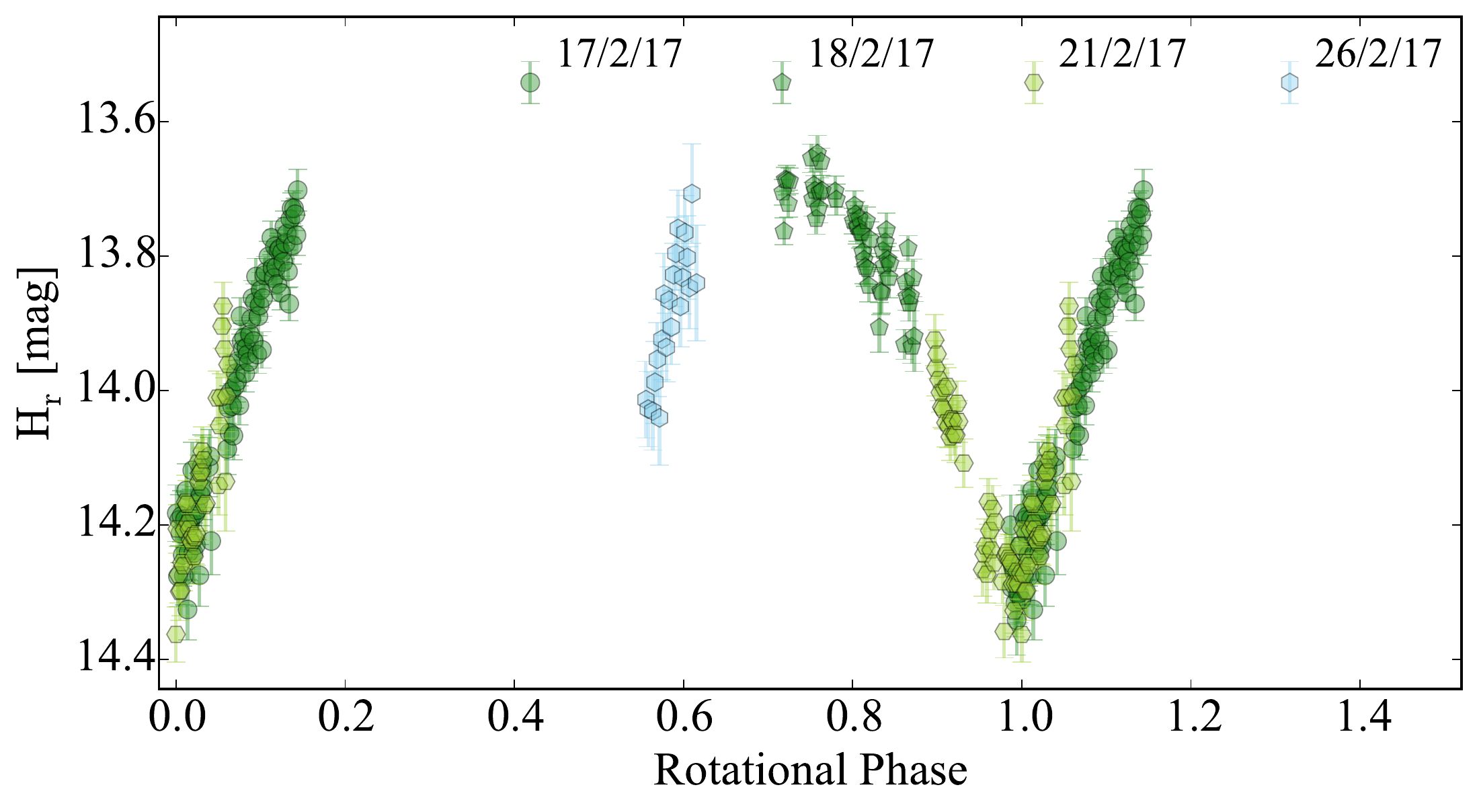}
   \caption{Rotational lightcurve of comet 162P from the data taken in February 2017, corrected for a phase-function slope $\beta$ = 0.039 mag/deg. The lightcurve is phased with $P$ = 32.92 hours. The magnitudes from 17 and 18 February 2017 were calibrated using the same set of comparison stars, and are therefore plotted in the same colour.}
    \label{162P_LC_NEW}%
    \end{figure}

The LS periodogram in Fig. \ref{162P_NEW_LS} has a maximum corresponding to a double-peaked lightcurve with $P$ = 32.92 hours. The corresponding lightcurve is plotted in Fig. \ref{162P_LC_NEW}. Due to the long rotation period of the comet, the observations from the INT only covered one of the lightcurve minima. However, due to the very dense sampling of the data close to the pronounced V-shaped minimum, a relatively narrow range of periods is able to produce a good alignment between the points from the different nights during the INT run. 

\begin{figure}
    \centering
  \includegraphics[width=0.48\textwidth]{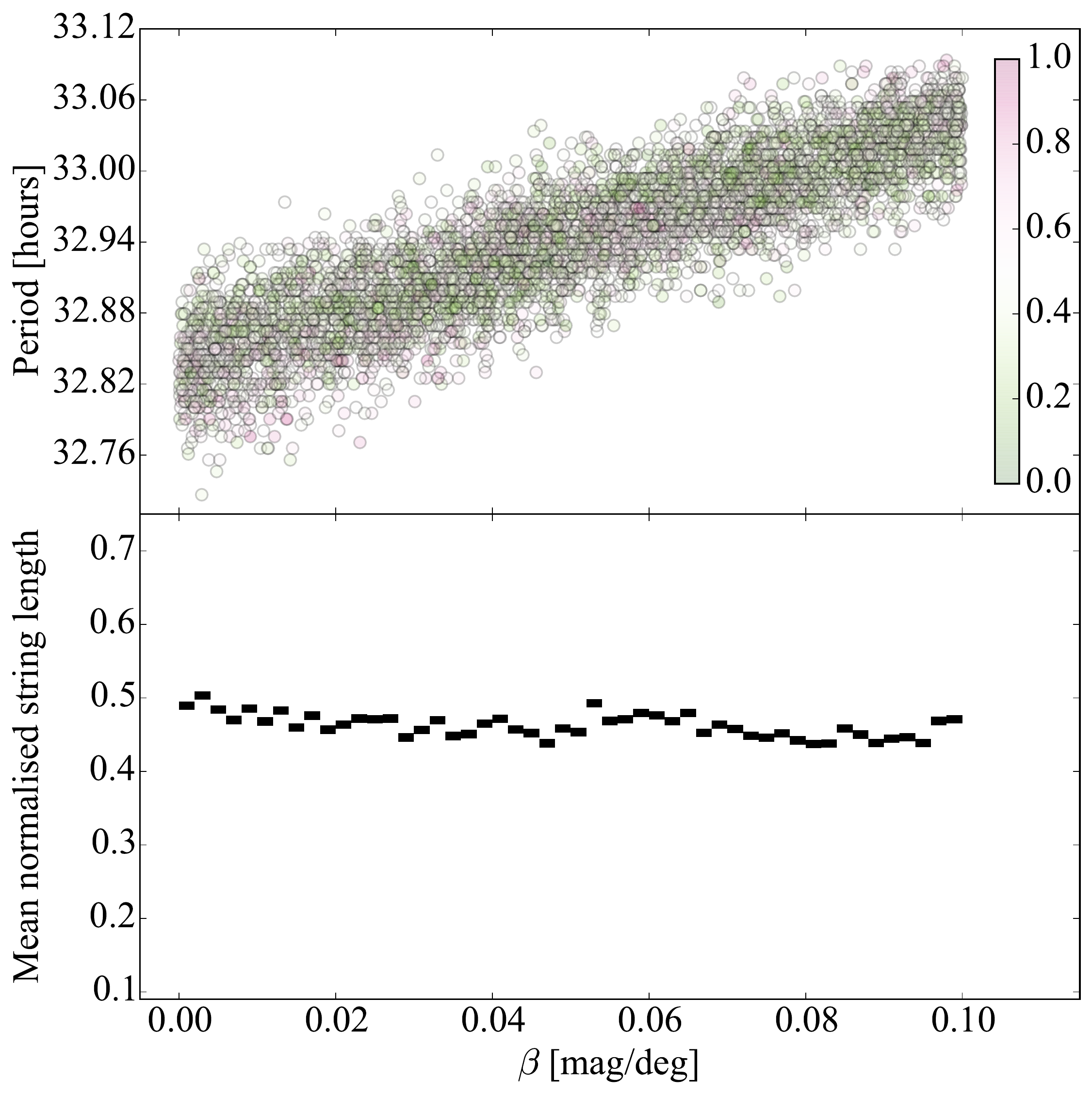}
   \caption{Same as Fig. \ref{14P_2016_MC2} for the 162P data taken in 2017. The MC2 method was run for phase-function slopes in the range 0.00 - 0.10 mag/deg and periods from 3 to 60 hours.}
    \label{162P_NEW_MC}%
    \end{figure}

In order to determine the uncertainty of the period, we used the MC2 method for a broad range of phase-function slopes (0.0 - 0.1 mag/deg), and looked for periods in the range 3-60 hours. The results in Fig. \ref{162P_NEW_MC} confirmed that the exact rotation period is dependent on the adopted phase function, and that the probed phase angle range is too narrow to allow us to determine the phase function unambiguously from this data set. The possible rotation periods for the whole $\beta$-range lie between 32.72 and 33.09 hours. If we take the possible periods for $\beta$ = 0.039 $\pm$ 0.02 mag/deg, then the current rotation rate of comet 162P is in the range 32.83 - 33.00 hours.  


The range of possible rotation periods derived for the dataset taken in 2017 also includes the rotation period $P$ = 32.853 hours, which was previously derived as the best period for the combined dataset from 2007 and 2012 \citepalias{Kokotanekova2017}. This implies that the current dataset does not allow us to detect period changes for 162P between the three apparitions. We can, however, combine all datasets from all three apparitions and use the MC2 method to search for a common period.

\begin{figure}
    \centering
  \includegraphics[width=0.48\textwidth]{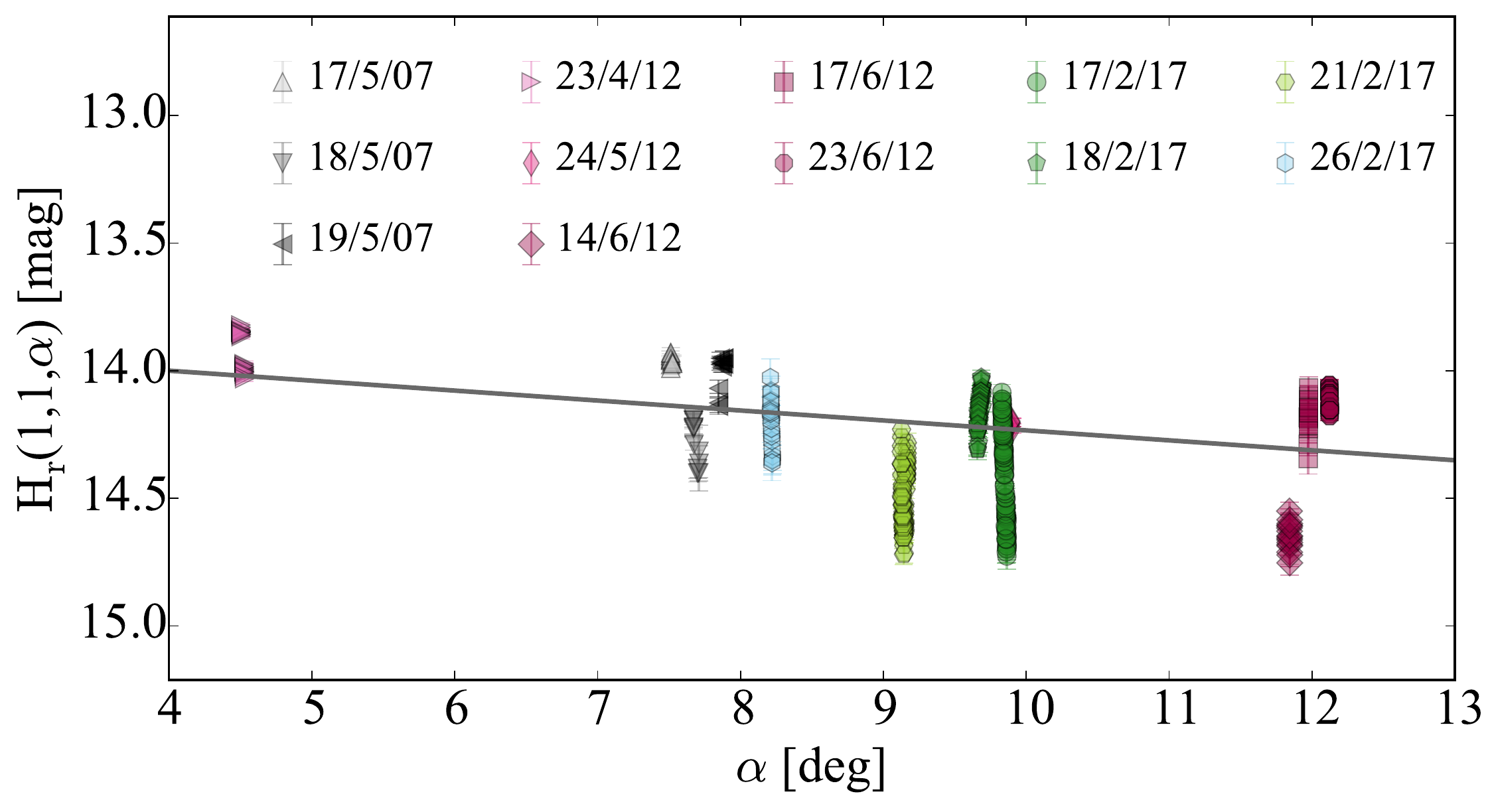}
   \caption{Phase function of comet 162P from the datasets taken in 2007, 2012 and 2017. The calibrated absolute magnitudes of the comet are plotted against phase angle $\alpha$. The magnitudes from 17 and 18 February 2017 were calibrated using the same set of comparison stars, and are therefore plotted in the same colour. Over-plotted is a linear phase function model with 0.039 mag/deg.}
    \label{162P_ALL_PHASE}%
    \end{figure}

In Fig. \ref{162P_ALL_PHASE} we have plotted the phase function of the combined dataset from all three epochs.  A linear fit to all points results in a phase-function slope $\beta$ = 0.035 mag/deg. The phase-function slope $\beta$ = 0.039 mag/deg from \citetalias{Kokotanekova2017} also produces a good fit to the data. The phase function is well-sampled at phase angles between 7 and 12 degrees, but the only observations outside of this range are a short dataset at $\alpha$ $\sim$ 4.7 degrees from April 2012. Due to the long period of the comet and the large brightness variation, even this extended dataset does not allow an unambiguous direct determination of the phase function. 

Since we were unable to determine the exact value of the phase-function slope from a direct fit, we ran the MC2 method for the full range of possible phase functions -  between 0.0 and 0.1 mag/deg. We looked for possible periods in the range 32.7 - 33.1 hours, which we determined above. 

\begin{figure}
  \centering
  \includegraphics[width=0.48\textwidth]{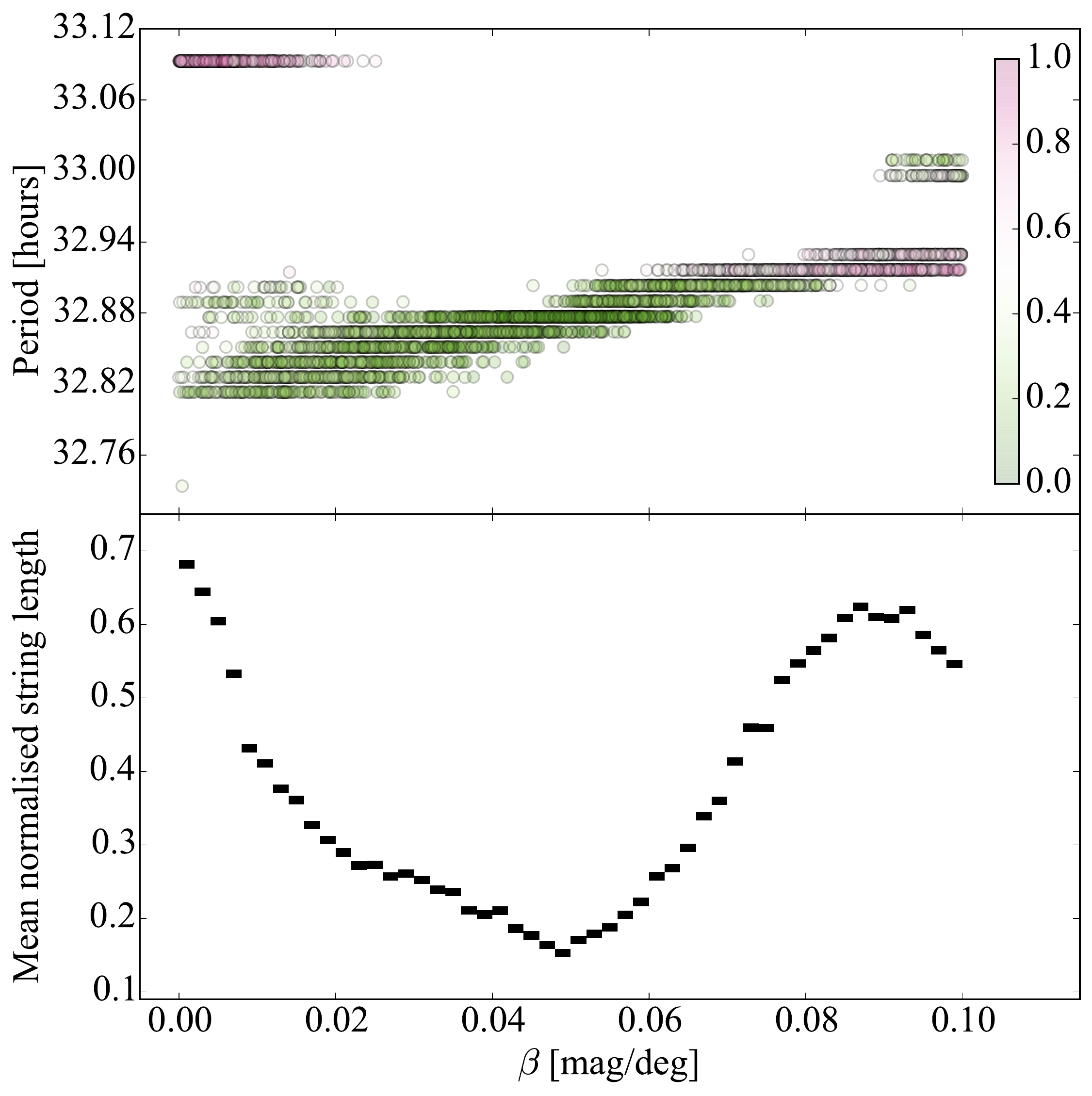}
   \caption{Same as Fig. \ref{14P_2016_MC2} for 162P from the combined datasets taken in 2007, 2012 and 2017. The MC2 method was run for a range of possible phase-function slopes $\beta$ = 0.00 - 0.10 mag/deg and periods from 32.7 to 33.1 hours.  }
    \label{162P_ALL_MC}%
    \end{figure}

\begin{figure}
   \centering
  \includegraphics[width=0.48\textwidth]{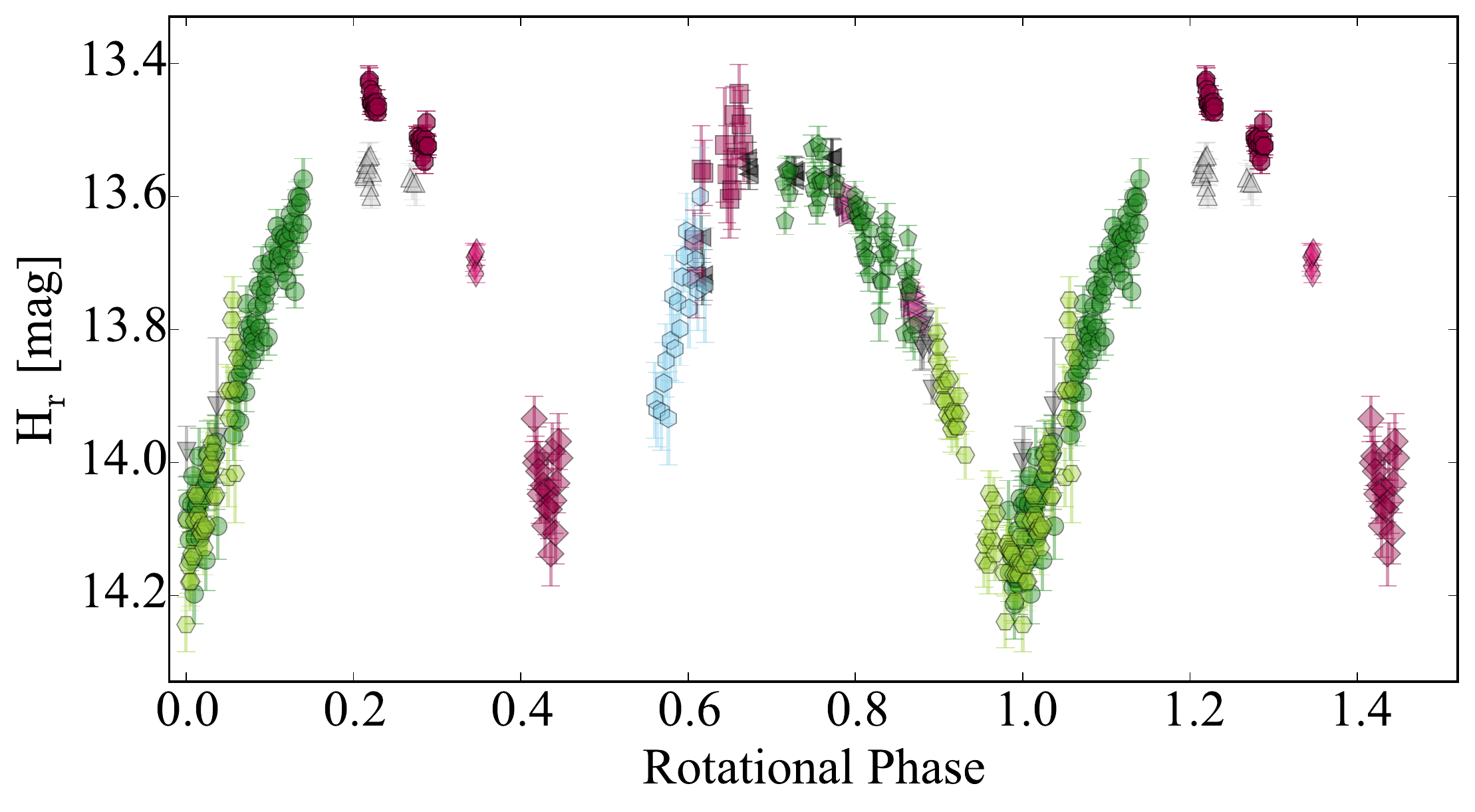}
   \caption{Rotational lightcurve of comet 162P with the combined dataset from 2007, 2012 and 2017. The symbols correspond to these in Fig. \ref{162P_ALL_PHASE}. The points were corrected for a phase-function slope $\beta$ = 0.052 mag/deg and phased with a rotation period $P$ = 32.877 hours.}
    \label{162P_LC_ALL}%
    \end{figure}

Fig. \ref{162P_ALL_MC} displays the results of the MC2 test. The best lightcurves were found for phase-function slopes of approximately 0.05 mag/deg and rotation rates of 32.877 hours. To illustrate the results, we have plotted the lightcurve of 162P from one of the combinations of $\beta$ and period which produced the best lightcurves in the MC2 test (Fig. \ref{162P_LC_ALL}). This lightcurve is representative for the best solutions from the MC2 test and illustrates the very good alignment between the individual datasets. 

We visually inspected the lightcurves of the clones with periods 32.73, 33.0--33.1 and 32.91--32.93 hours and confirmed that they show poor agreement with the data. We therefore conclude that the range of possible common periods for the datasets from 2007, 2012 and 2017 is 32.812--32.903 hours. Additional observations during the current aphelion arc may allow this to be refined further, in order to search for subtle changes in future orbits.

The common lightcurve with the data from all three apparitions shows a good match between the peak width and brightness variation of the individual datasets. There is a small offset between the points from 2007 and 2012 at rotational phase $\sim$ 0.2. The possible differences in peak height from the different apparitions could be due to changing viewing geometry. However the overall agreement between the three datasets implies that it is possible to find a common rotation period for all epochs. We therefore have no evidence that there was a period change between the three epochs. However, to set a formal upper limit on the spin change we take the difference between the maximum possible period for 2012 \citepalias[33.237 hours;][]{Kokotanekova2017} and the minimum period for 2017, 32.83 to derive a conservative upper limit of 25 min in the past orbit.


\section{Discussion}
\label{sec:discussion}

\begin{table*}
\centering
\caption{Properties of all JFCs with observed period changes.}
\label{table_changes}
\begin{tabular}{lllll}
\hline
Name & Radius (km) & Period (hours) & Period change (min/orbit) & References \\
\hline
14P/Wolf & 2.95	$\pm$ 0.19 & 9 & $<$ 4.2 & (1), (2), this paper \\
143P/K-M & $\mathrm{4.79^{+0.32}_{-0.33}}$ & 17 & $<$ 6.6 & (1), (3), this paper \\
162P/S-S & $\mathrm{7.03^{+0.47}_{-0.48}}$ & 33 & $<$ 25 & (1), (2), this paper \\
\hline
2P/Encke & 3.95 $\pm$ 0.06 & 11 & 4 & (4), (5) \\
9P/Tempel 1 & 2.83 $\pm$ 0.1 & 41 & -13.49 & (6), (7), (8) \\
10P/Tempel 2 & 5.98	$\pm$ 0.04 & 9 & 0.27 & (9), (10), (11), (12), (13) \\
19P/Borelly & 2.5	$\pm$ 0.1 & 29 & 20 & (14), (15), (16) \\
41P/T-G-K & 0.7-1 & 20 & $>$ 1560\textsuperscript{a} & (17), (18) \\
49P/A-R & 4.24 $\pm$ 0.2 & 13 & $<$ 0.23 & (19), (20), (21), (22) \\
67P/C-G & 1.649 $\pm$ 0.007 & 12 & -20.95 & (23), ESA/Rosetta \\
103P/Hartley 2 & 0.58	$\pm$ 0.018 & 16 & 120 & (24),  (25), (26), (27)\\
\hline
\end{tabular}
\begin{minipage}{\textwidth}
	\textsuperscript{a} The period change of more than 26 hours for comet 41P was measured during the same apparition. \\
	References: (1) \cite{Fernandez2013}; (2) \cite{Kokotanekova2017}; (3) \cite{Jewitt2003}; (4) \cite{Lowry2007}; (5) \cite{Samarasinha2013}; (6) \cite{Thomas2013}; (7) \cite{Belton2011}; (8) \cite{Chesley2013}; (9) \cite{Lamy2009}; (10) \cite{Mueller1996}; (11) \cite{Knight2011}; (12) \cite{Knight2012}; (13) \cite{Schleicher2013}; (14) \cite{Buratti2004}; (15) \cite{Mueller2002}; (16) \cite{Mueller2015}; (17) \cite{Tancredi2000}; (18) \cite{Bodewits2018}; (19) \cite{Lamy2004}; (20) \cite{Millis1988}; (21) \cite{Campins1995}; (22) \cite{Eisner2017}; (23) \cite{Jorda2016}; (24) \cite{Thomas2013a}; (25) \cite{Meech2009}; (26) \cite{Meech2011b}; (27) \cite{Jehin2010}; \\
\end{minipage}
\end{table*}

\subsection{Survivability of large JFC nuclei}
\label{sec:disc_survivability}

In this work, we compared newly obtained photometric observations of three large JFCs (14P, 143P and 162P) to their previous lightcurves from past orbits. For each of the three comets we were able to find a common period which describes well the combined data from the different apparitions. Even though this strongly suggests that the comets did not experience significant period changes, due to the uncertainties in the previous lightcurves and the phase functions, we have chosen to place conservative upper limits on the spin changes. 

\begin{figure}
   \centering
  \includegraphics[width=0.48\textwidth]{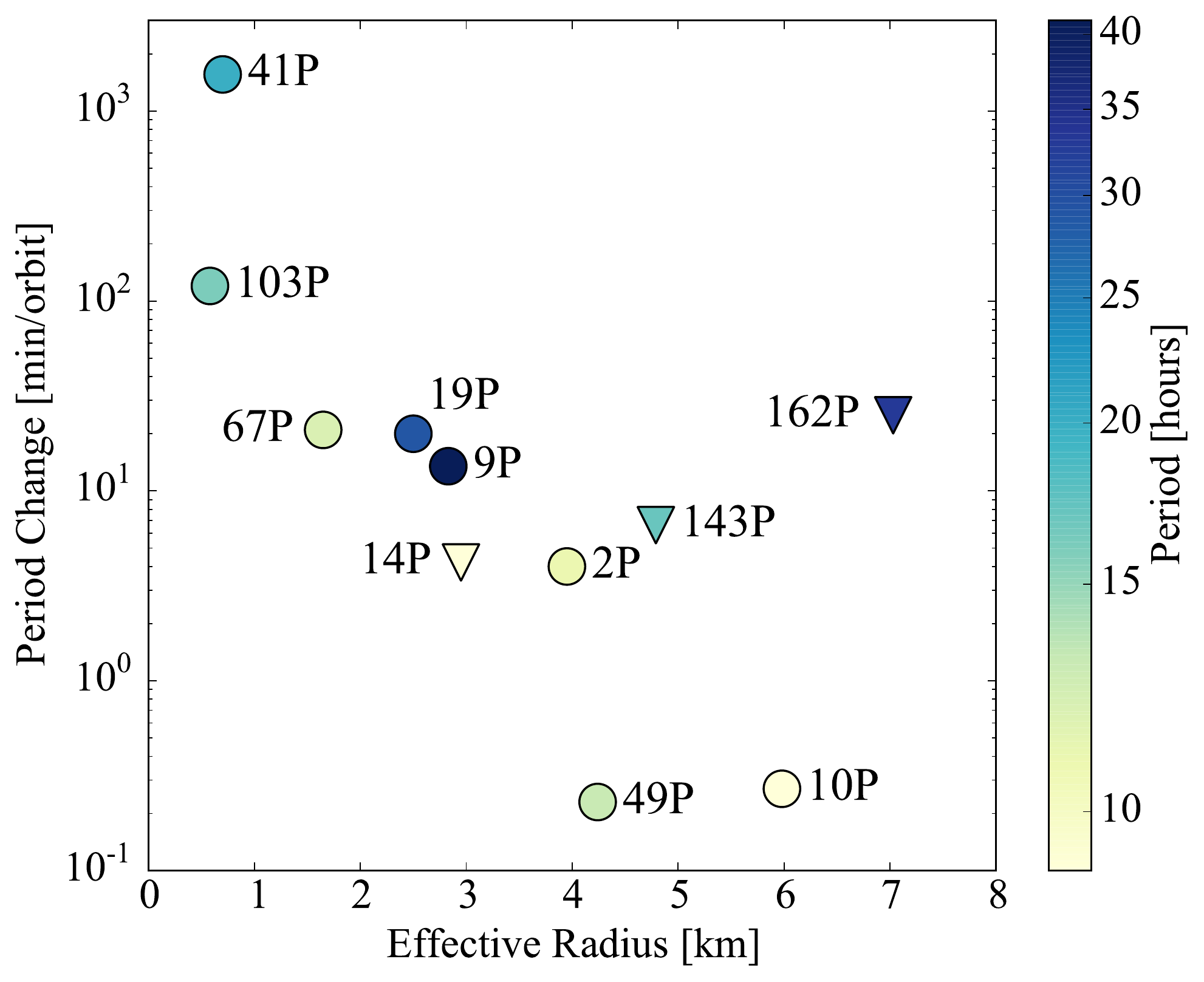}
   \caption{Comparison between the JFCs nuclei with known period changes. The circles show comets from the literature. The triangles correspond to the upper limits for comets from this work. The colours of the points correspond to the rotation period of the comets. The two smallest nuclei, 41P and 103P have displayed the most noticeable period changes of 26 and 2 hours respectively. On the contrary, the largest nuclei exhibit the smallest period changes.}
    \label{changes_summary_plot}%
    \end{figure}

In Table \ref{table_changes} and Fig. \ref{changes_summary_plot}, we compare the parameters of the three comets from this work to all other JFCs with detected spin changes. Prior to this work, spin changes were measured for eight other JFCs. It is noticeable that the two smallest nuclei, 103P and 41P, displayed the largest period changes, of $\sim$2 hours per orbit \citep{Meech2011b} and $>$26 hours per orbit \citep{Bodewits2018}, respectively. The three comets with sizes in the range 1-3 km had period changes of the order of tens of minutes, while the three largest nuclei, 2P, 10P and 49P, had $\Delta P$ $<$ 10 min. 

The three comets analysed in this work have $R$ $\geq$ 3 km and belong to the largest JFCs. Therefore the non-detection of spin changes is in agreement with the observations of the other large JFCs. For comets 14P and 143P, the conservative upper limits derived in Section \ref{sec:results} also match the expected period changes $\Delta P$ $<$ 10 minutes.

    
The observed trend of decreasing period change with increasing radius is predicted by simple theoretical considerations of the changing spin rate due to outgassing. For instance, according to \cite{Samarasinha2013}, for comets with similar densities, shapes and activity distributions, the period changes decrease for increasing effective radii and decreasing rotation periods (faster rotation). It is also expected that comets with lower levels of outgassing will experience smaller period changes. 

In \citetalias{Kokotanekova2017} we noticed that JFCs with $R$ $\geq$ 3 km lie well above the rotational-instability limit derived for the whole population of JFCs. We then hypothesised that this is due to the small period changes these comets are expected to undergo given their large radii. With the current work, we have added small upper limits for the period changes of three comets in this size range. These findings confirm the prediction that large JFCs experience very small spin-rate changes, and are not expected to reach the rotational instability limit.

Out of the comets with $R$ $\geq$ 3 km in Table \ref{table_changes}, 2P has a moderate activity level while all other comets can be described as very weakly active (see \cite{Jewitt2003}, \cite{Samarasinha2013}, \cite{Eisner2017}, \citetalias{Kokotanekova2017} and references therein). Having both large sizes and low activity levels makes these comets less likely to experience significant activity-driven period changes. They are therefore also less likely to undergo activity-induced rotational splitting, and more likely than smaller and more active comets to survive more perihelion passages without significant mass loss. 

It may be possible for weakly active and dormant comets to experience an enhancement in activity without changing their orbits. If this happens, then the long-term stability of these objects might be disturbed. For example, motivated by the fly-by observations of comet 103P, \cite{Steckloff2016} suggested that a relatively fast nucleus rotation can cause avalanches which are able to expose fresh volatile-rich material and to reactivate previously dormant comets. This scenario, however, requires that the comet spins up to reach a necessary minimum rotation rate to trigger this event. Considering the small period changes discovered for the large JFCs discussed above, it seems improbable that they would be affected by this reactivation mechanism. This once again suggests that if their orbits remain stable, larger nuclei will most likely remain weakly active or dormant, and will therefore survive longer than smaller comets.

We have identified three further lines of observational evidence which are in favour of the idea that larger JFCs have an increased survivability. Firstly, \cite{Fernandez2013} identified a bump in the cumulative size distribution (CSD) of JFCs for effective radii between 3 and 6 km. This implies an excess of large nuclei. However, since the number of comets that fall into this size range is small, this observation needs to be considered with caution. In order to confirm its validity and to verify whether the excess is just for radii of 3-6 km or extends to larger nuclei, it is necessary to increase the number of JFCs with precisely measured sizes. 

Secondly, recent works on the CSD of dead comets in the ACO population \citep{Kim2014,Licandro2016} report a flatter cumulative size distribution for dormant comets than for active JFCs. Provided that the selection criteria of these two studies successfully distinguish between asteroids and dormant/dead comets, and that this finding is not a result of observational bias towards preferentially observing larger objects \cite[see the discussion in][]{Kim2014}, the flatter CSD slope implies that the larger nuclei preferentially survive the active phase of their evolution compared to smaller comets. 

Finally, dynamical studies following the orbital evolution of small bodies incoming from the Kuiper Belt fail to reproduce the observed distribution of short-period comets \citep{DiSisto2009,Rickman2017,Nesvorny2017}. The discrepancies between the numerical models and observations, however, can be reduced significantly if a different physical lifetime for comets of different sizes is introduced. In  particular \cite{Nesvorny2017} made an estimate that 10-km-class comets should survive thousands of perihelion passages while 1-km-class comets should only survive on the order of hundreds of perihelion passages, and 100-meter-sized nuclei should only live for a few perihelion passages.

In addition to the decreased likelihood for a spin-up and rotationally-driven instability, there are further mechanisms that could contribute to increase the survivability of large JFCs and can be evoked to explain these findings. Generally, ground observations have suggested that large JFC nuclei are often characterized by low levels of activity \citep[e.g.][]{AHearn1995,Tancredi2006}. This tendency is explained with a variety of models that involve the formation of devolatised dust mantles which prevent the sublimation of the underlying material and can eventually make the comet dormant or dead \citep[see][]{Jewitt2004a}. The observations of dust deposits on comet 67P by Rosetta's OSIRIS cameras have confirmed that some large particles are unable to leave the comet's gravitational field and get redeposited on the nucleus surface \citep{Thomas2015}. Following this idea, the larger the comets, the stronger their gravitational potential, and therefore the more particles will get trapped in their gravitational field and will eventually return to the nucleus. Thus, larger nuclei will build insulating layers after fewer perihelion passages and will become dormant before they could undergo large mass loss. 

\cite{Gundlach2016} have proposed an alternative model to explain why the CSDs of JFCs and ACOs differ for objects with radii $>$ 2 km \citep{Kim2014}. \cite{Gundlach2016} suggest that the interiors of bigger nuclei have experienced larger hydrostatic compression and as a result have a larger tensile strength. At a certain point when the activity-driven erosion of the comet reaches deeper, more compacted layers and the sublimation is no longer able to lift off the dust particles from the surface, the activity of the comet ceases \citep{Gundlach2016}. Hence, this mechanism also implies that larger nuclei become inactive after fewer perihelion passages.	   

In both scenarios, since large nuclei become inactive faster than smaller ones, they are more likely to preserve their large sizes during the evolution as active comets. Provided that their average heliocentric distances remain unchanged over time, large JFCs remain shielded by their surface layers and are also less likely to undergo large mass-loss events (outbursts and splitting). 

In summary, all of the outlined mechanisms imply that the combined effects of the larger size and the low activity of JFCs with effective radii larger than 2-3 km makes them more resistant to rotational splitting and other processes responsible for significant mass loss in comets. We can therefore conclude that large JFC nuclei must have an enhanced survivability with respect to their smaller counterparts.

\subsection{Surface evolution of JFC nuclei}
\label{sec:disc_surface}

   \begin{figure}
   \centering
  \includegraphics[width=0.48\textwidth]{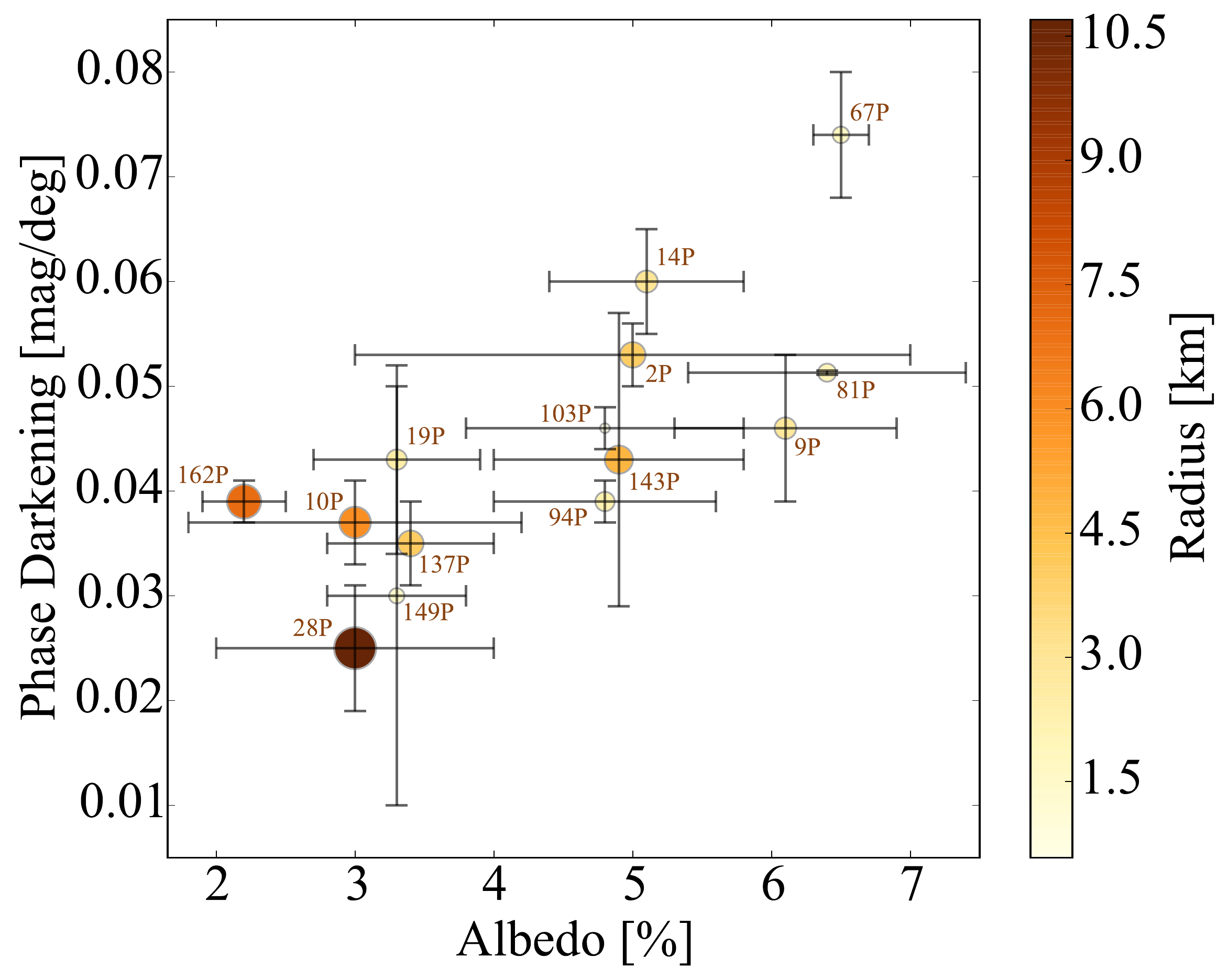}
   \caption{Linear phase-function slope $\beta$ versus geometric albedo in R-band for all JFCs with measurements of both parameters. The size of the symbols and their colours correspond to the effective radii of the nuclei. The albedo for 143P was derived in this work while all other values are taken from Table \ref{tab:surface}. Despite the large uncertainties in the measurements, the distribution of the comets in this plot suggests a correlation between the phase-function slope and the albedo. The largest and least active nuclei appear to be clustered at lower $\beta$ and albedo.} 
    \label{surface_param}%
    \end{figure}
    
Using the newly obtained optical observations of comet 143P and its radius estimate from \cite{Fernandez2013}, we derived a geometric albedo $p_{\mathrm{r_{P1}}}$ = 0.048 $\pm$ 0.009. This allowed us to add 143P to the small set of 13 comets from \citetalias{Kokotanekova2017} which have reliable albedo and phase function estimates (Table \ref{tab:surface}). In Fig. \ref{surface_param} we have plotted the linear phase-function slopes $\beta$ versus the geometric albedos in R-band for all 14 comets in this set. In \citetalias{Kokotanekova2017} we discussed that the largest JFCs nuclei have comparatively smaller albedos and less steep phase functions. We also identified a possible correlation between the linear phase-function slope and geometric albedo. Comet 143P agrees with the observed trend and appears to have moderate albedo and phase-function slope. 

\begin{table*}
\centering
\caption{Properties of all JFCs with known albedos and phase functions slopes.}
\label{tab:surface}
\begin{tabular}{llllllll}
\hline
Comet & $p_{\mathrm{R}}$ [\%]\textsuperscript{*} & Reference                                             & $\beta$ [mag/deg]   & Range & Reference                                          &  Radius [km]                             & Reference            \\
\hline
2P    & 5.0 $\pm$ 2.0       & (1)    						& 0.053 $\pm$ 0.003		& 0-110 	& Weighted mean (1,2) & 3.95 $\pm$ 0.06     & (3)    \\
9P    & 6.1 $\pm$ 0.8       & Weighted mean (4,5,6)	& 0.046 $\pm$ 0.007   	& 4-117 	& (4)                    & 2.83 $\pm$ 0.1      & (7)    \\
10P   & 3.0 $\pm$ 1.2       & (8)                           & 0.037 $\pm$ 0.004   	& 9-28  	& (9)                    & 5.98 $\pm$ 0.04     & (10)	\\
14P   & 5.1 $\pm$ 0.7       & (11)                         	& 0.060 $\pm$ 0.005   	& 5-9   	& (11)                   & 2.95 $\pm$ 0.19     & (12)	\\
19P   & 3.3 $\pm$ 0.6       & Weighted mean (13,14)      & 0.043 $\pm$ 0.009   	& 13-80 	& (13)                   & 2.5 $\pm$ 0.1       & (14)   \\
28P   & 3.0 $\pm$ 1.0       & (15)                          & 0.025 $\pm$ 0.006   	& 0-15  	& (16)                   & 10.7 $\pm$ 0.7      & (17)   \\
67P   & 6.5 $\pm$ 0.2       & (18)                          & 0.074 $\pm$ 0.006   	& 1-10  	& (18)                   & 1.649 $\pm$ 0.007   & (19)   \\
81P   & 6.4 $\pm$ 1.0       & (20)                          & 0.0513 $\pm$ 0.0002 	& 0-100 	& (20)                   & 1.98 $\pm$ 0.05     & (21)  	\\
94P   & 4.8 $\pm$ 0.8       & (11)                         	& 0.039 $\pm$ 0.002   	& 5-17  	& (11)                   & $\mathrm{2.27^{0.13}_{0.15}}$ & (12) \\
103P  & 4.8 $\pm$ 1.0       & (22)                          & 0.046 $\pm$ 0.002   	& 79-95 	& (22)                   & 0.58 $\pm$ 0.018    & (23)   \\
137P  & 3.4 $\pm$ 0.6       & (11)                         	& 0.035 $\pm$ 0.004   	& 0.5-6 	& (11)                   & $\mathrm{4.04^{0.31}_{0.32}}$ & (12) \\
143P  & 4.9 $\pm$ 0.9       & This work                     & 0.043 $\pm$ 0.014   	& 5-13  	& (24)                   & $\mathrm{4.79^{0.32}_{0.33}}$ & (12) \\
149P  & 3.3 $\pm$ 0.5       & (11)                         	& 0.03 $\pm$ 0.02     	& 8-10  	& (11)                   & $\mathrm{1.42^{0.09}_{0.10}}$ & (12) \\
162P  & 2.2 $\pm$ 0.3       & (11)                         	& 0.039 $\pm$ 0.002   	& 4-12  	& (11)                   & $\mathrm{7.03^{0.47}_{0.48}}$ & (12) \\
\hline
\end{tabular}
\begin{minipage}{\textwidth}
\textsuperscript{*} Albedos are in R-band, converted from $\mathrm{r_{P1}}$ where necessary. The conversion is done using $p_{\mathrm{R}}$ = $p_{\mathrm{r_{P1}}}$ $\times$ 1.021 for the mean colour index (B-V) = 0.87 $\pm$ 0.05 mag \citep{Lamy2009a}.\\
References: (1) \cite{Fernandez2000}, (2) \citep{Boehnhardt2008}, (3) \cite{Lowry2007}, (4) \cite{Li2007b}, (5) \cite{Lisse2005}, (6) \cite{Fernandez2003}, (7) \cite{Thomas2013}, (8) \cite{AHearn1989}, (9) \cite{Sekanina1991}, (10) \cite{Lamy2009}, (11) \citetalias{Kokotanekova2017}, (12) \cite{Fernandez2013}, (13) \cite{Li2007a}, (14) \cite{Buratti2004}, (15) \cite{Jewitt1988}, (16) \cite{Delahodde2001}, (17) \cite{Lamy2004}, (18) \cite{Fornasier2015a}, (19) \cite{Jorda2016}, (20) \cite{Li2009},(21) \cite{Sekanina2004},(22) \cite{Li2013},(23) \cite{Thomas2013a},(24) \cite{Jewitt2003}
\end{minipage}
\end{table*}

Before we proceed to discuss the possible interpretation of the phase function-albedo correlation, we emphasise that it is based on a small set of comets. Moreover, the error bars in Fig. \ref{surface_param} clearly indicate the large uncertainties associated with each measurement. Even the measurements of comets 9P \citep{Li2007b}, 19P \citep{Li2007a}, 67P \citep{Fornasier2015a}, 81P \citep{Li2009} and 103P \citep{Li2013} made during spacecraft visits have large uncertainties, which highlights the technical difficulties intrinsic to photometric studies of cometary surfaces. Since it is unlikely that observations in the near future will allow the uncertainties of the albedo and the phase-function slopes to be decreased, the best way to verify the validity of the correlation is to increase the number of comets in the diagram with future ground observations. 

We also note that the phase functions for the different comets were measured for different $\alpha$ ranges. Even though the Rosetta observations allowed the detection of an opposition surge of comet 67P \citep{Fornasier2015a,Masoumzadeh2017,Hasselmann2017}, the opposition effect was not observed during the fly-bys of other comets, or in any ground-based measurement to date. This suggests that linear fits provide a good approximation to the phase functions, and hence the slopes derived from phase-function observations of different $\alpha$ ranges must be comparable. 

Keeping in mind these possible caveats, we proceed to interpret the trend in Fig. \ref{surface_param} in light of the recent in-situ studies of cometary surfaces. There is now an increasing body of evidence that the surface morphology and texture of comet nuclei is governed by sublimation-driven erosion and that it reflects the degree of evolution of the comets \citep[e.g.][]{Basilevsky2006,Ip2016,Vincent2017}. Moreover, the different surface morphologies are believed to produce detectable differences in the comets' optical properties \citep[e.g.][]{Fornasier2015a,Longobardo2017}.

After a comparison of the three comets visited by spacecraft at the time, \cite{Basilevsky2006} noticed that smooth flat surfaces become more prevalent in the sequence 81P, 9P, 19P. They accounted this to progressive sublimational degradation, which increases with the number of perihelion passages. 

During the Rosetta visit to 67P, \cite{Ip2016} investigated whether the size frequency distribution of circular depressions of the different comets could be related to their dynamical history. They performed orbital integration simulations which showed that comets 67P, 103P and 19P could have spent more time orbiting at heliocentric distances under 2.5 au, thus being more eroded than 81P and 9P. It is however necessary to point out that such dynamical studies are complicated by the non-gravitational forces caused by outgassing and by the chaotic nature of JFC orbits which can vary greatly depending on the initial conditions of the orbital integration. Therefore, the suggested evolution sequence has to be taken with caution. In particular, it is not certain how recently 67P has entered the inner Solar System, and it is possible that it has experienced less erosion than 103P and 19P \citep[see][]{Ip2016,Vincent2017}.  

The most comprehensive evidence for the connection between the surface morphology and the erosion levels of JFCs comes from \cite{Vincent2017}. They compared the cumulative cliff-height distribution on different regions of 67P and of three other comets visited by spacecraft, 9P, 81P and 103P. They discovered that the regions on comet 67P which receive the highest insolation are lacking large cliffs. \cite{Vincent2017} hypothesised that instead of simply losing mass due to sublimation, comet nuclei, whose topography is initially dominated by steep cliffs, gradually get eroded down to flatter surfaces composed of smaller fragments (pebbles and dust). 

The comparison between 67P and the other nuclei imaged during spacecraft fly-bys is in agreement with the proposed mechanism \citep{Vincent2017}. The power index of the cumulative cliff height distribution decreases in the order 81P, 67P, 9P, 103P, suggesting that the level of erosion of these comets increases in this direction \citep{Vincent2017}. This sequence is generally supported by the findings of the dynamical studies of \cite{Ip2016}, once more implying that the global surface morphology can be related to the level of erosion of the nucleus. 

The different surface morphologies, on the other hand, can be related to different photometric behaviour. \cite{Longobardo2017} used the VIRTIS imaging spectrometer on board Rosetta and discovered that rougher terrains on 67P produce slightly steeper phase functions. They also concluded that comets 81P and 9P, which have rougher surfaces, are photometrically similar to C-type asteroids and have phase functions steeper than those of smoother comets (103P, 19P and 67P). Using the orbital evolution studies by \cite{Ip2016}, they suggested that comets which have experienced more sublimation-driven erosion have smoother surfaces and less steep phase functions. 

All of these studies motivated us to look for a connection between the phase function-albedo correlation in Fig. \ref{surface_param} and the level of surface erosion of the individual comets. Comets 81P and 9P, which should have experienced more surface erosion according to \cite{Ip2016}, indeed have larger albedos and phase-function slopes than 19P and 103P, which should be dynamically younger (although it is hard to distinguish 103P from 9P due to their large uncertainties). 

Comet 67P is the one with the highest albedo and highest phase-function slope. However, according to \cite{Ip2016} it should not be the least eroded nucleus among those visited by spacecraft. We account this discrepancy to the fact that the albedo and phase-function slope in Fig. \ref{surface_param} are taken from \cite{Fornasier2015a}, and were obtained before perihelion when only the northern hemisphere of the nucleus was observable. Due to the rotational axis orientation of 67P, the northern hemisphere of the nucleus receives less insolation throughout the orbit, and is therefore less eroded than the southern hemisphere \citep{Keller2015,Vincent2017}. It is therefore very likely that the southern hemisphere would have a smaller phase-function slope and albedo. However, to our knowledge no direct comparison between the optical properties of the two hemispheres is available at the time of writing this paper. 

Finally, at the bottom left corner of the plot in Fig. \ref{surface_param}, at low albedos and flat phase functions, we can find three of the largest JFCs - 10P, 28P and 162P. Comet 10P is known to have weak activity at perihelion, while 28P and 162P have very weak and intermittent activity and have been classified as transition objects on the way to become dead comets \citep{AHearn1995,Campins2006}.

\subsection{Evolution hypothesis}

Considering all of the evidence presented above, we propose the following hypothesis to explain the correlation between $\beta$ and geometric albedo: Dynamically young JFCs begin their lives as active comets having volatile-rich and rough surfaces characterised by tall steep cliffs. These surfaces correspond to relatively high albedos of 6-7 \% and steep phase functions with slopes $\beta$ $>$ 0.04 mag/deg. As the comets orbit around the Sun, their primitive topography gets gradually eroded and  gives place to smoother terrains, which correspond to flatter phase functions. Towards the end of their lives as active comets, the nuclei are covered by ever-growing dust areas which gradually quench the activity. As they gradually transition to dormant comets, the volatiles from the surface layers gradually sublimate, which results in a further albedo decrease.

As we discussed in Section \ref{sec:disc_survivability}, the larger nuclei are less susceptible to major mass-loss mechanisms (splitting/disruption), and are therefore more likely to reach a state of complete surface erosion. Hence, finding the large and almost dead comets at the bottom left corner of Fig. \ref{surface_param} supports our hypothesis. 

Interestingly, some of the highest albedos and phase-function slopes are found for the comets visited by spacecraft (9P, 67P and 81P). This raises the question whether there is a discrepancy between values derived from ground observations and from modelling disc-resolved photometry from spacecraft data. It must be considered, however, that space-mission teams aimed to select targets with well-known orbits and well-characterised behaviour. These criteria were satisfied mainly by comets which were discovered early on due to their high activity and the larger brightness corresponding to it. Therefore, it is understandable why the surfaces of more evolved and less active comets have remained unobserved by space missions. A future mission visiting a low-activity or dormant comet would be very interesting for comparison.

The majority of the comets in Fig. \ref{surface_param} were observed with ground- and space-based telescopes \citepalias[see][]{Kokotanekova2017}. Therefore, the possible phase function-albedo correlation provides a compelling opportunity to study the surface characteristics and evolution of JFCs from the ground. Moreover this correlation could provide us with the possibility to distinguish between asteroids which have been placed on cometary orbits and dormant/dead comets. If the correlation is true, then dead comets which have undergone full erosion will have lower albedos and flatter phase functions than those of C-type asteroids. 

These prospects emphasise the need to confirm and  better understand the observed trends in the photometric properties of JFCs. This can be achieved by: 
\begin{enumerate}
\item{Increasing the sample of JFCs with well-constrained geometric albedos and phase functions from ground-based observations;}

\item{Performing thorough dynamical studies of the orbital history of all comets with known surface characteristics;}

\item{Comparing the observed phase functions to laboratory samples in order to understand the material properties behind the observed albedos and phase functions; }

\item{Understanding the effects of large-scale topography on the phase functions; }

\item{Comparing the photometric properties of JFCs with those of Centaurs and Kuiper Belt objects, which should be similar to less-eroded comets; }

\item{Comparing the photometric properties of JFCs and asteroids on cometary orbits.} 

\end{enumerate}

\section{Summary}
\label{sec:conclusions}

We have collected photometric time-series observations for three large JFCs, 14P, 143P and 162P, in order to derive their current rotation periods and to look for changes with respect to their spin rates from previous apparitions. We determined the following periods from the new lightcurves: $P$ = 9.07 $\pm$ 0.01 hours for 14P; $P_1$ = 17.1966 $\pm$ 0.0003 hours, $P_2$ = 17.2121 $\pm$ 0.0002 hours or $P_3$ = 17.1812 $\pm$ 0.0002 hours for 143P; $P$ = 32.9 $\pm$ 0.2 hours for 162P. For each of the three comets we were able to find a common period which phases well all previously published lightcurves. Thus, we were unable to detect spin changes with respect to the last apparitions but we set conservative upper limits for the spin change of $\Delta P$ < 4.2 min per orbit (14P), $\Delta P$ < 6.6 min per orbit (143P) and $\Delta P$ < 25 min per orbit (162P). 

With the new observations we have increased the number of JFCs with studied period changes from eight to eleven. This expanded sample shows clear evidence that the largest JFC nuclei with $R$ $\geq$ 3 km experience the smallest period changes (typically $\Delta P$ < 10 minutes). This observation implies that large comets are less likely to undergo significant period changes and rotational splitting over their lifetimes. We have also discussed other processes which can contribute to prevent large JFCs from undergoing significant mass-loss events. This led to the conclusion that the interplay of all mechanisms makes nuclei of large JFCs more likely to survive their evolution as active comets until they reach full surface erosion and transition to dormancy. The suggested enhanced survivability of large JFCs can explain the CSD of JFCs from \cite{Fernandez2013} and of dormant comets in the ACO population from \cite{Kim2014} and \cite{Licandro2016}, all of which have suggested an excess of objects with radii larger than 2.5 - 3 km. 

Our new observations of comet 143P allowed us to derive a geometric albedo $p_{\mathrm{r_{P1}}}$ = 0.048	$\pm$ 0.009. We added it to the small sample of JFCs with well-constrained phase functions and geometric albedos from \citetalias{Kokotanekova2017}. The 14 comets in Fig. \ref{surface_param} follow a trend of increasing phase-function slope with increasing albedo. 

In light of recent detailed studies of the surfaces of JFCs visited by spacecraft, we hypothesise that this possible correlation corresponds to an evolutionary trend for JFCs. In this scenario, dynamically young JFCs start their evolution with relatively high albedos and steeper phase functions. During their lifetime as active JFCs, sublimation-driven erosion gradually makes their surfaces smoother and their phase-function slopes decrease. As the dust-covered portions of the nuclei progressively increase, the comets become less active and the sublimation gradually decreases. Finally, the dust layers gradually lose their volatiles and therefore their albedos decrease even further as the comets transition to dormancy. 

If confirmed, this trend in the photometric parameters offers a fascinating opportunity to study the evolution of cometary surfaces with ground-based observations. It could also provide a criterion to distinguish cometary bodies from asteroids on comet-like orbits.



\section*{Acknowledgements}

We would like to thank Jean-Baptiste Vincent for helpful discussions about the surface evolution of comets.

RK acknowledges financial support of the Research Group in Cometary Science and the International Max Planck Research School for Solar System Science at Max Planck Institute for Solar System Research in G\"ottingen, Germany. CS is funded by a STFC Ernest Rutherford fellowship. SFG acknowledges the financial support of the UK STFC (grants ST/L000776/1 and ST/P000657/1). 

PN gratefully acknowledge the financial support from the project DFNP 106/11.05.2016 - "Program for career development of young scientists", Bulgarian Academy of Sciences (BAS).  The research of PN and TB were partially supported by the following contracts DN 18/13-12.12.2017 and DN 08-1/2016 with the Bulgarian National Science Fund.

We thank the observatory staff who helped us obtain the various data sets in this paper. 
Based on observations made with the INT operated on the island of La Palma by the Isaac Newton Group of Telescopes in the Spanish Observatorio del Roque de los Muchachos of the Instituto de Astrofisica de Canarias, under UK PATT programme I/2017A/05. 
Based on observations collected at the German-Spanish Astronomical Center, Calar Alto, jointly operated by the Max-Planck-Institut f\"ur Astronomie Heidelberg and the Instituto de Astrof\'isica de Andaluc\'ia (CSIC).

The Pan-STARRS1 Surveys (PS1) and the PS1 public science archive have been made possible through contributions by the Institute for Astronomy, the University of Hawaii, the Pan-STARRS Project Office, the Max-Planck Society and its participating institutes, the Max Planck Institute for Astronomy, Heidelberg and the Max Planck Institute for Extraterrestrial Physics, Garching, The Johns Hopkins University, Durham University, the University of Edinburgh, the Queen's University Belfast, the Harvard-Smithsonian Center for Astrophysics, the Las Cumbres Observatory Global Telescope Network Incorporated, the National Central University of Taiwan, the Space Telescope Science Institute, the National Aeronautics and Space Administration under Grant No. NNX08AR22G issued through the Planetary Science Division of the NASA Science Mission Directorate, the National Science Foundation Grant No. AST-1238877, the University of Maryland, Eotvos Lorand University (ELTE), the Los Alamos National Laboratory, and the Gordon and Betty Moore Foundation.




\bibliographystyle{mnras}
\bibliography{Comets_bib.bib} 

\begin{thebibliography}{}
\makeatletter
\relax
\def\mn@urlcharsother{\let\do\@makeother \do\$\do\&\do\#\do\^\do\_\do\%\do\~}
\def\mn@doi{\begingroup\mn@urlcharsother \@ifnextchar [ {\mn@doi@}
  {\mn@doi@[]}}
\def\mn@doi@[#1]#2{\def\@tempa{#1}\ifx\@tempa\@empty \href
  {http://dx.doi.org/#2} {doi:#2}\else \href {http://dx.doi.org/#2} {#1}\fi
  \endgroup}
\def\mn@eprint#1#2{\mn@eprint@#1:#2::\@nil}
\def\mn@eprint@arXiv#1{\href {http://arxiv.org/abs/#1} {{\tt arXiv:#1}}}
\def\mn@eprint@dblp#1{\href {http://dblp.uni-trier.de/rec/bibtex/#1.xml}
  {dblp:#1}}
\def\mn@eprint@#1:#2:#3:#4\@nil{\def\@tempa {#1}\def\@tempb {#2}\def\@tempc
  {#3}\ifx \@tempc \@empty \let \@tempc \@tempb \let \@tempb \@tempa \fi \ifx
  \@tempb \@empty \def\@tempb {arXiv}\fi \@ifundefined
  {mn@eprint@\@tempb}{\@tempb:\@tempc}{\expandafter \expandafter \csname
  mn@eprint@\@tempb\endcsname \expandafter{\@tempc}}}

\bibitem[\protect\citeauthoryear{A'Hearn, Campins, Schleicher  \&
  Millis}{A'Hearn et~al.}{1989}]{AHearn1989}
A'Hearn M.~F.,  Campins H.,  Schleicher D.~G.,   Millis R.~L.,  1989, \mn@doi
  [The Astrophysical Journal] {10.1086/168204}, 347, 1155

\bibitem[\protect\citeauthoryear{A'Hearn, Millis, Schleicher, Osip  \&
  Birch}{A'Hearn et~al.}{1995}]{AHearn1995}
A'Hearn M.~F.,  Millis R.,  Schleicher D.,  Osip D.,   Birch P.,  1995, \mn@doi
  [Icarus] {10.1006/icar.1995.1190}, 118, 223

\bibitem[\protect\citeauthoryear{Basilevsky \& Keller}{Basilevsky \&
  Keller}{2006}]{Basilevsky2006}
Basilevsky A.,  Keller H.,  2006, \mn@doi [Planetary and Space Science]
  {10.1016/j.pss.2006.05.001}, 54, 808

\bibitem[\protect\citeauthoryear{Belton et~al.,}{Belton
  et~al.}{2011}]{Belton2011}
Belton M.~J.,  et~al., 2011, \mn@doi [Icarus] {10.1016/j.icarus.2011.01.006},
  213, 345

\bibitem[\protect\citeauthoryear{Bodewits, Farnham, Kelley  \& Knight}{Bodewits
  et~al.}{2018}]{Bodewits2018}
Bodewits D.,  Farnham T.~L.,  Kelley M. S.~P.,   Knight M.~M.,  2018, \mn@doi
  [Nature] {10.1038/nature25150}, 553, 186

\bibitem[\protect\citeauthoryear{Boehnhardt}{Boehnhardt}{2004}]{Boehnhardt2004}
Boehnhardt H.,  2004, Comets II

\bibitem[\protect\citeauthoryear{Boehnhardt, Tozzi, Bagnulo, Muinonen, Nathues
  \& Kolokolova}{Boehnhardt et~al.}{2008}]{Boehnhardt2008}
Boehnhardt H.,  Tozzi G.~P.,  Bagnulo S.,  Muinonen K.,  Nathues A.,
  Kolokolova L.,  2008, \mn@doi [Astronomy and Astrophysics]
  {10.1051/0004-6361:200809922}, 489, 1337

\bibitem[\protect\citeauthoryear{Buratti, Hicks, Soderblom, Britt, Oberst  \&
  Hillier}{Buratti et~al.}{2004}]{Buratti2004}
Buratti B.,  Hicks M.,  Soderblom L.,  Britt D.,  Oberst J.,   Hillier J.,
  2004, \mn@doi [Icarus] {10.1016/j.icarus.2003.05.002}, 167, 16

\bibitem[\protect\citeauthoryear{Campins, Osip, Rieke  \& Rieke}{Campins
  et~al.}{1995}]{Campins1995}
Campins H.,  Osip D.~J.,  Rieke G.,   Rieke M.,  1995, \mn@doi [Planetary and
  Space Science] {10.1016/0032-0633(94)E0074-Z}, 43, 733

\bibitem[\protect\citeauthoryear{Campins, Ziffer, Licandro, Pinilla-Alonso,
  Fern{\'{a}}ndez, de Le{\'{o}}n, Moth{\'{e}}-Diniz  \& Binzel}{Campins
  et~al.}{2006}]{Campins2006}
Campins H.,  Ziffer J.,  Licandro J.,  Pinilla-Alonso N.,  Fern{\'{a}}ndez Y.,
  de Le{\'{o}}n J.,  Moth{\'{e}}-Diniz T.,   Binzel R.~P.,  2006, \mn@doi [The
  Astronomical Journal] {10.1086/506253}, 132, 1346

\bibitem[\protect\citeauthoryear{Chambers et~al.,}{Chambers
  et~al.}{2016}]{Chambers2016}
Chambers K.~C.,  et~al., 2016, http://arxiv.org/abs/1612.05560

\bibitem[\protect\citeauthoryear{Chesley et~al.,}{Chesley
  et~al.}{2013}]{Chesley2013}
Chesley S.,  et~al., 2013, \mn@doi [Icarus] {10.1016/j.icarus.2012.03.022},
  222, 516

\bibitem[\protect\citeauthoryear{Davidsson}{Davidsson}{1999}]{Davidsson1999}
Davidsson B. J.~R.,  1999, \mn@doi [Icarus] {10.1006/icar.1999.6214}, 142, 525

\bibitem[\protect\citeauthoryear{Davidsson}{Davidsson}{2001}]{Davidsson2001}
Davidsson B. J.~R.,  2001, \mn@doi [Icarus] {10.1006/icar.2000.6540}, 149, 375

\bibitem[\protect\citeauthoryear{Davidsson, Sierks, Guttler, Marzari, Pajola
  \& Rickman}{Davidsson et~al.}{2016}]{Davidsson2016}
Davidsson B. J.~R.,  Sierks H.,  Guttler C.,  Marzari F.,  Pajola M.,   Rickman
  H.,  2016, \mn@doi [Astronomy {\&} Astrophysics]
  {10.1051/0004-6361/201526968}, 592, A63

\bibitem[\protect\citeauthoryear{Delahodde, Meech, Hainaut  \& Dotto}{Delahodde
  et~al.}{2001}]{Delahodde2001}
Delahodde C.~E.,  Meech K.~J.,  Hainaut O.~R.,   Dotto E.,  2001, \mn@doi
  [Astronomy and Astrophysics] {10.1051/0004-6361:20011028}, 376, 672

\bibitem[\protect\citeauthoryear{{Di Sisto}, Fern{\'{a}}ndez  \& Brunini}{{Di
  Sisto} et~al.}{2009}]{DiSisto2009}
{Di Sisto} R.~P.,  Fern{\'{a}}ndez J.~A.,   Brunini A.,  2009, \mn@doi [Icarus]
  {10.1016/j.icarus.2009.05.002}, 203, 140

\bibitem[\protect\citeauthoryear{Duncan \& Levison}{Duncan \&
  Levison}{1997}]{Duncan1997a}
Duncan M.~J.,  Levison H.~F.,  1997, \mn@doi [Science]
  {10.1126/science.276.5319.1670}, 276, 1670

\bibitem[\protect\citeauthoryear{Dworetsky}{Dworetsky}{1983}]{Dworetsky1983}
Dworetsky M.~M.,  1983, Monthly Notices of the Royal Astronomical Society, 203,
  917

\bibitem[\protect\citeauthoryear{Eisner, Knight  \& Schleicher}{Eisner
  et~al.}{2017}]{Eisner2017}
Eisner N.,  Knight M.~M.,   Schleicher D.~G.,  2017, \mn@doi [The Astronomical
  Journal] {10.3847/1538-3881/aa8b0b}, 154, 196

\bibitem[\protect\citeauthoryear{Fern{\'{a}}ndez}{Fern{\'{a}}ndez}{2000}]{Fernandez2000}
Fern{\'{a}}ndez Y.,  2000, \mn@doi [Icarus] {10.1006/icar.2000.6431}, 147, 145

\bibitem[\protect\citeauthoryear{Fern{\'{a}}ndez, Jewitt  \&
  Sheppard}{Fern{\'{a}}ndez et~al.}{2001}]{Fernandez2001}
Fern{\'{a}}ndez Y.~R.,  Jewitt D.~C.,   Sheppard S.~S.,  2001, \mn@doi [The
  Astrophysical Journal] {10.1086/320689}, 553, L197

\bibitem[\protect\citeauthoryear{Fern{\'{a}}ndez, Meech, Lisse, A'Hearn,
  Pittichov{\'{a}}  \& Belton}{Fern{\'{a}}ndez et~al.}{2003}]{Fernandez2003}
Fern{\'{a}}ndez Y.,  Meech K.,  Lisse C.,  A'Hearn M.,  Pittichov{\'{a}} J.,
  Belton M.,  2003, \mn@doi [Icarus] {10.1016/S0019-1035(03)00142-8}, 164, 481

\bibitem[\protect\citeauthoryear{Fern{\'{a}}ndez, Jewitt  \&
  Sheppard}{Fern{\'{a}}ndez et~al.}{2005}]{Fernandez2005a}
Fern{\'{a}}ndez Y.~R.,  Jewitt D.~C.,   Sheppard S.~S.,  2005, \mn@doi [The
  Astronomical Journal] {10.1086/430802}, 130, 308

\bibitem[\protect\citeauthoryear{Fern{\'{a}}ndez et~al.,}{Fern{\'{a}}ndez
  et~al.}{2013}]{Fernandez2013}
Fern{\'{a}}ndez Y.,  et~al., 2013, \mn@doi [Icarus]
  {10.1016/j.icarus.2013.07.021}, 226, 1138

\bibitem[\protect\citeauthoryear{Fornasier et~al.,}{Fornasier
  et~al.}{2015}]{Fornasier2015a}
Fornasier S.,  et~al., 2015, \mn@doi [Astronomy {\&} Astrophysics]
  {10.1051/0004-6361/201525901}, 583, A30

\bibitem[\protect\citeauthoryear{Gundlach, Blum  \& Blum}{Gundlach
  et~al.}{2016}]{Gundlach2016}
Gundlach B.,  Blum J.,   Blum J.,  2016, \mn@doi [Astronomy {\&} Astrophysics]
  {10.1051/0004-6361/201527260}, 589, A111

\bibitem[\protect\citeauthoryear{Guti{\'{e}}rrez, Jorda, Samarasinha  \&
  Lamy}{Guti{\'{e}}rrez et~al.}{2005}]{Gutierrez2005}
Guti{\'{e}}rrez P.~J.,  Jorda L.,  Samarasinha N.~H.,   Lamy P.,  2005, \mn@doi
  [Planetary and Space Science] {10.1016/j.pss.2004.12.012}, 53, 1135

\bibitem[\protect\citeauthoryear{Harris, Young, Scaltriti  \& Zappalà}{Harris
  et~al.}{1984}]{Harris1984}
Harris A.,  Young J.,  Scaltriti F.,   Zappalà V.,  1984, \mn@doi [Icarus]
  {10.1016/0019-1035(84)90070-8}, 57, 251–258

\bibitem[\protect\citeauthoryear{Hasselmann et~al.,}{Hasselmann
  et~al.}{2017}]{Hasselmann2017}
Hasselmann P.~H.,  et~al., 2017, \mn@doi [Monthly Notices of the Royal
  Astronomical Society] {10.1093/mnras/stx1834}, 469, S550

\bibitem[\protect\citeauthoryear{Hodgkin, Irwin, Hewett  \& Warren}{Hodgkin
  et~al.}{2008}]{Hodgkin2008}
Hodgkin S.~T.,  Irwin M.~J.,  Hewett P.~C.,   Warren S.~J.,  2008, \mn@doi
  [Monthly Notices of the Royal Astronomical Society]
  {10.1111/j.1365-2966.2008.14387.x}, 394, 675

\bibitem[\protect\citeauthoryear{Ip et~al.,}{Ip et~al.}{2016}]{Ip2016}
Ip W.-H.,  et~al., 2016, \mn@doi [Astronomy {\&} Astrophysics]
  {10.1051/0004-6361/201628156}, 591, A132

\bibitem[\protect\citeauthoryear{Jehin, Manfroid, Hutsemekers, Gillon  \&
  Magain}{Jehin et~al.}{2010}]{Jehin2010}
Jehin E.,  Manfroid J.,  Hutsemekers D.,  Gillon M.,   Magain P.,  2010,
  Central Bureau Electronic Telegrams, 2589

\bibitem[\protect\citeauthoryear{Jewitt}{Jewitt}{2004}]{Jewitt2004a}
Jewitt D.~C.,  2004, in , Comets II.
pp 659--676

\bibitem[\protect\citeauthoryear{Jewitt \& Meech}{Jewitt \&
  Meech}{1988}]{Jewitt1988}
Jewitt D.~C.,  Meech K.~J.,  1988, \mn@doi [The Astrophysical Journal]
  {10.1086/166351}, 328, 974

\bibitem[\protect\citeauthoryear{Jewitt, Sheppard  \& Fernndez}{Jewitt
  et~al.}{2003}]{Jewitt2003}
Jewitt D.,  Sheppard S.,   Fernndez Y.,  2003, \mn@doi [The Astronomical
  Journal] {10.1086/374947}, 125, 3366

\bibitem[\protect\citeauthoryear{Jorda et~al.,}{Jorda et~al.}{2016}]{Jorda2016}
Jorda L.,  et~al., 2016, \mn@doi [Icarus] {10.1016/j.icarus.2016.05.002}, 277,
  257

\bibitem[\protect\citeauthoryear{Kaiser et~al.,}{Kaiser
  et~al.}{2002}]{Kaiser2002}
Kaiser N.,  et~al., 2002, in Survey and Other Telescope Technologies and
  Discoveries. Proceedings of the SPIE. p.~154

\bibitem[\protect\citeauthoryear{Kaiser et~al.,}{Kaiser
  et~al.}{2010}]{Kaiser2010}
Kaiser N.,  et~al., 2010, in Ground-based and Airborne Telescopes III.
  Proceedings of the SPIE. p. 77330E

\bibitem[\protect\citeauthoryear{Keller, Mottola, Skorov  \& Jorda}{Keller
  et~al.}{2015}]{Keller2015}
Keller H.~U.,  Mottola S.,  Skorov Y.,   Jorda L.,  2015, \mn@doi [Astronomy
  {\&} Astrophysics] {10.1051/0004-6361/201526421}, 579, L5

\bibitem[\protect\citeauthoryear{Kim, Ishiguro  \& Usui}{Kim
  et~al.}{2014}]{Kim2014}
Kim Y.,  Ishiguro M.,   Usui F.,  2014, \mn@doi [The Astrophysical Journal]
  {10.1088/0004-637X/789/2/151}, 789, 151

\bibitem[\protect\citeauthoryear{Knight, Farnham, Schleicher  \&
  Schwieterman}{Knight et~al.}{2011}]{Knight2011}
Knight M.~M.,  Farnham T.~L.,  Schleicher D.~G.,   Schwieterman E.~W.,  2011,
  \mn@doi [The Astronomical Journal] {10.1088/0004-6256/141/1/2}, 141, 2

\bibitem[\protect\citeauthoryear{Knight, Schleicher, Farnham, Schwieterman  \&
  Christensen}{Knight et~al.}{2012}]{Knight2012}
Knight M.~M.,  Schleicher D.~G.,  Farnham T.~L.,  Schwieterman E.~W.,
  Christensen S.~R.,  2012, \mn@doi [The Astronomical Journal]
  {10.1088/0004-6256/144/5/153}, 144, 153

\bibitem[\protect\citeauthoryear{Kokotanekova et~al.,}{Kokotanekova
  et~al.}{2017}]{Kokotanekova2017}
Kokotanekova R.,  et~al., 2017, \mn@doi [Monthly Notices of the Royal
  Astronomical Society] {10.1093/mnras/stx1716}, 471, 2974

\bibitem[\protect\citeauthoryear{Lamy \& Toth}{Lamy \& Toth}{2009}]{Lamy2009a}
Lamy P.,  Toth I.,  2009, \mn@doi [Icarus] {10.1016/j.icarus.2009.01.030}, 201,
  674

\bibitem[\protect\citeauthoryear{Lamy, Toth, Fernandez  \& Weaver}{Lamy
  et~al.}{2004}]{Lamy2004}
Lamy P.~L.,  Toth I.,  Fernandez Y.~R.,   Weaver H.~A.,  2004, in , Comets II.
pp 223--264

\bibitem[\protect\citeauthoryear{Lamy, Toth, Weaver, A'Hearn  \& Jorda}{Lamy
  et~al.}{2009}]{Lamy2009}
Lamy P.~L.,  Toth I.,  Weaver H.~A.,  A'Hearn M.~F.,   Jorda L.,  2009, \mn@doi
  [Astronomy and Astrophysics] {10.1051/0004-6361/200811462}, 508, 1045

\bibitem[\protect\citeauthoryear{Levison \& Duncan}{Levison \&
  Duncan}{1997}]{Levison1997}
Levison H.,  Duncan M.~J.,  1997, \mn@doi [Icarus] {10.1006/icar.1996.5637},
  127, 13

\bibitem[\protect\citeauthoryear{Levison, Morbidelli, Vanlaerhoven, Gomes  \&
  Tsiganis}{Levison et~al.}{2008}]{LEVISON2008}
Levison H.,  Morbidelli A.,  Vanlaerhoven C.,  Gomes R.,   Tsiganis K.,  2008,
  \mn@doi [Icarus] {10.1016/j.icarus.2007.11.035}, 196, 258

\bibitem[\protect\citeauthoryear{Li et~al.,}{Li et~al.}{2007a}]{Li2007b}
Li J.-Y.,  et~al., 2007a, \mn@doi [Icarus] {10.1016/j.icarus.2006.09.018}, 187,
  41

\bibitem[\protect\citeauthoryear{Li, A'Hearn, McFadden  \& Belton}{Li
  et~al.}{2007b}]{Li2007a}
Li J.,  A'Hearn M.,  McFadden L.,   Belton M.,  2007b, \mn@doi [Icarus]
  {10.1016/j.icarus.2006.11.015}, 188, 195

\bibitem[\protect\citeauthoryear{Li, A'Hearn, Farnham  \& McFadden}{Li
  et~al.}{2009}]{Li2009}
Li J.-Y.,  A'Hearn M.~F.,  Farnham T.~L.,   McFadden L.~A.,  2009, \mn@doi
  [Icarus] {10.1016/j.icarus.2009.06.002}, 204, 209

\bibitem[\protect\citeauthoryear{Li et~al.,}{Li et~al.}{2013}]{Li2013}
Li J.-Y.,  et~al., 2013, \mn@doi [Icarus] {10.1016/j.icarus.2012.11.001}, 222,
  559

\bibitem[\protect\citeauthoryear{Licandro, Ali-Lagoa, Tancredi  \&
  Fernandez}{Licandro et~al.}{2016}]{Licandro2016}
Licandro J.,  Ali-Lagoa V.,  Tancredi G.,   Fernandez Y.,  2016, \mn@doi
  [Astronomy {\&} Astrophysics] {10.1051/0004-6361/201526866}, 585, A9

\bibitem[\protect\citeauthoryear{Lisse et~al.,}{Lisse et~al.}{2005}]{Lisse2005}
Lisse C.~M.,  et~al., 2005, \mn@doi [The Astrophysical Journal]
  {10.1086/431238}, 625, L139

\bibitem[\protect\citeauthoryear{Lomb}{Lomb}{1976}]{Lomb1976}
Lomb N.~R.,  1976, \mn@doi [Astrophysics and Space Science]
  {10.1007/BF00648343}, 39, 447

\bibitem[\protect\citeauthoryear{Longobardo et~al.,}{Longobardo
  et~al.}{2017}]{Longobardo2017}
Longobardo A.,  et~al., 2017, \mn@doi [Monthly Notices of the Royal
  Astronomical Society] {10.1093/mnras/stx1803}, 469, S346

\bibitem[\protect\citeauthoryear{Lowry \& Weissman}{Lowry \&
  Weissman}{2007}]{Lowry2007}
Lowry S.~C.,  Weissman P.~R.,  2007, \mn@doi [Icarus]
  {10.1016/j.icarus.2006.11.014}, 188, 212

\bibitem[\protect\citeauthoryear{Masoumzadeh et~al.,}{Masoumzadeh
  et~al.}{2017}]{Masoumzadeh2017}
Masoumzadeh N.,  et~al., 2017, \mn@doi [Astronomy {\&} Astrophysics]
  {10.1051/0004-6361/201629734}, 599, A11

\bibitem[\protect\citeauthoryear{Meech et~al.,}{Meech et~al.}{2009}]{Meech2009}
Meech K.~J.,  et~al., 2009, in American Astronomical Society, DPS meeting
  {\#}41, id.20.07.

\bibitem[\protect\citeauthoryear{Meech et~al.,}{Meech
  et~al.}{2011}]{Meech2011b}
Meech K.~J.,  et~al., 2011, \mn@doi [The Astrophysical Journal]
  {10.1088/2041-8205/734/1/L1}, 734, L1

\bibitem[\protect\citeauthoryear{Millis, A'Hearn  \& Campins}{Millis
  et~al.}{1988}]{Millis1988}
Millis R.~L.,  A'Hearn M.~F.,   Campins H.,  1988, \mn@doi [The Astrophysical
  Journal] {10.1086/165974}, 324, 1194

\bibitem[\protect\citeauthoryear{Mueller \& Ferrin}{Mueller \&
  Ferrin}{1996}]{Mueller1996}
Mueller B.~E.,  Ferrin I.,  1996, \mn@doi [Icarus] {10.1006/icar.1996.0172},
  123, 463

\bibitem[\protect\citeauthoryear{Mueller \& Samarasinha}{Mueller \&
  Samarasinha}{2002}]{Mueller2002}
Mueller B. E.~A.,  Samarasinha N.~H.,  2002, Earth, Moon, and Planets, 90, 463

\bibitem[\protect\citeauthoryear{Mueller \& Samarasinha}{Mueller \&
  Samarasinha}{2015}]{Mueller2015}
Mueller B. E.~A.,  Samarasinha N.~H.,  2015, in American Astronomical Society,
  DPS meeting {\#}47 id.506.10.

\bibitem[\protect\citeauthoryear{Nesvorny, Vokrouhlicky, Dones, Levison, Kaib
  \& Morbidelli}{Nesvorny et~al.}{2017}]{Nesvorny2017}
Nesvorny D.,  Vokrouhlicky D.,  Dones L.,  Levison H.~F.,  Kaib N.,
  Morbidelli A.,  2017, \mn@doi [The Astrophysical Journal]
  {10.3847/1538-4357/aa7cf6}, 845, 27

\bibitem[\protect\citeauthoryear{Pravec, Sarounova  \& Wolf}{Pravec
  et~al.}{1996}]{Pravec1996}
Pravec P.,  Sarounova L.,   Wolf M.,  1996, \mn@doi [Icarus]
  {10.1006/icar.1996.0223}, 124, 471

\bibitem[\protect\citeauthoryear{Rickman, Gabryszewski, Wajer,
  Wi{\'{s}}niowski, W{\'{o}}jcikowski, Szutowicz, Valsecchi  \&
  Morbidelli}{Rickman et~al.}{2017}]{Rickman2017}
Rickman H.,  Gabryszewski R.,  Wajer P.,  Wi{\'{s}}niowski T.,
  W{\'{o}}jcikowski K.,  Szutowicz S.,  Valsecchi G.~B.,   Morbidelli A.,
  2017, \mn@doi [Astronomy {\&} Astrophysics] {10.1051/0004-6361/201629374},
  598, A110

\bibitem[\protect\citeauthoryear{Samarasinha \& Mueller}{Samarasinha \&
  Mueller}{2013}]{Samarasinha2013}
Samarasinha N.~H.,  Mueller B. E.~A.,  2013, \mn@doi [The Astrophysical
  Journal] {10.1088/2041-8205/775/1/L10}, 775, L10

\bibitem[\protect\citeauthoryear{Samarasinha, Mueller, Belton  \&
  Jorda}{Samarasinha et~al.}{2004}]{Samarasinha2004}
Samarasinha N.~H.,  Mueller B. E.~A.,  Belton M. J.~S.,   Jorda L.,  2004, in ,
  Comets II.
pp 281--299

\bibitem[\protect\citeauthoryear{Scargle}{Scargle}{1982}]{Scargle1982}
Scargle J.~D.,  1982, \mn@doi [The Astrophysical Journal] {10.1086/160554},
  263, 835

\bibitem[\protect\citeauthoryear{Schleicher, Knight  \& Levine}{Schleicher
  et~al.}{2013}]{Schleicher2013}
Schleicher D.~G.,  Knight M.~M.,   Levine S.~E.,  2013, \mn@doi [The
  Astronomical Journal] {10.1088/0004-6256/146/5/137}, 146, 137

\bibitem[\protect\citeauthoryear{Sekanina \& Zdenek}{Sekanina \&
  Zdenek}{1991}]{Sekanina1991}
Sekanina Z.,  Zdenek 1991, \mn@doi [The Astronomical Journal] {10.1086/115881},
  102, 350

\bibitem[\protect\citeauthoryear{Sekanina, Brownlee, Economou, Tuzzolino  \&
  Green}{Sekanina et~al.}{2004}]{Sekanina2004}
Sekanina Z.,  Brownlee D.~E.,  Economou T.~E.,  Tuzzolino A.~J.,   Green S.~F.,
   2004, \mn@doi [Science] {10.1126/science.1098388}, 304, 1769

\bibitem[\protect\citeauthoryear{Snodgrass, Fitzsimmons  \& Lowry}{Snodgrass
  et~al.}{2005}]{Snodgrass2005}
Snodgrass C.,  Fitzsimmons A.,   Lowry S.~C.,  2005, \mn@doi [Astronomy and
  Astrophysics] {10.1051/0004-6361:20053237}, 444, 287

\bibitem[\protect\citeauthoryear{Steckloff, Graves, Hirabayashi, Melosh  \&
  Richardson}{Steckloff et~al.}{2016}]{Steckloff2016}
Steckloff J.~K.,  Graves K.,  Hirabayashi T.,  Melosh H.~J.,   Richardson J.,
  2016, \mn@doi [Icarus] {10.1016/j.icarus.2016.02.026}, 272, 60

\bibitem[\protect\citeauthoryear{Tancredi, Fern{\'{a}}ndez, Rickman  \&
  Licandro}{Tancredi et~al.}{2000}]{Tancredi2000}
Tancredi G.,  Fern{\'{a}}ndez J.~A.,  Rickman H.,   Licandro J.,  2000, \mn@doi
  [Astronomy and Astrophysics Supplement Series] {10.1051/aas:2000263}, 146, 73

\bibitem[\protect\citeauthoryear{Tancredi, Fern{\'{a}}ndez, Rickman  \&
  Licandro}{Tancredi et~al.}{2006}]{Tancredi2006}
Tancredi G.,  Fern{\'{a}}ndez J.~A.,  Rickman H.,   Licandro J.,  2006, \mn@doi
  [Icarus] {10.1016/j.icarus.2006.01.007}, 182, 527

\bibitem[\protect\citeauthoryear{Thomas et~al.,}{Thomas
  et~al.}{2013a}]{Thomas2013}
Thomas P.,  et~al., 2013a, \mn@doi [Icarus] {10.1016/j.icarus.2012.02.037},
  222, 453

\bibitem[\protect\citeauthoryear{Thomas et~al.,}{Thomas
  et~al.}{2013b}]{Thomas2013a}
Thomas P.~C.,  et~al., 2013b, \mn@doi [Icarus] {10.1016/j.icarus.2012.05.034},
  222, 550

\bibitem[\protect\citeauthoryear{Thomas et~al.,}{Thomas
  et~al.}{2015}]{Thomas2015}
Thomas N.,  et~al., 2015, \mn@doi [Astronomy {\&} Astrophysics]
  {10.1051/0004-6361/201526049}, 583, A17

\bibitem[\protect\citeauthoryear{Tody}{Tody}{1986}]{Tody1986}
Tody D.,  1986, in Crawford D.~L.,  ed.,  Vol. 0627, Instrumentation in
  astronomy VI; Proceedings of the Meeting, Tucson, AZ, Mar. 4-8, 1986. Part 2.
  p.~733

\bibitem[\protect\citeauthoryear{Tody}{Tody}{1993}]{Tody1993}
Tody D.,  1993, in Astronomical Data Analysis Software and Systems II, A.S.P.
  Conference Series. p.~173

\bibitem[\protect\citeauthoryear{Tonry et~al.,}{Tonry et~al.}{2012}]{Tonry2012}
Tonry J.~L.,  et~al., 2012, \mn@doi [The Astrophysical Journal]
  {10.1088/0004-637X/750/2/99}, 750, 99

\bibitem[\protect\citeauthoryear{Tsiganis, Gomes, Morbidelli  \&
  Levison}{Tsiganis et~al.}{2005}]{Tsiganis2005}
Tsiganis K.,  Gomes R.,  Morbidelli A.,   Levison H.~F.,  2005, \mn@doi
  [Nature] {10.1038/nature03539}, 435, 459

\bibitem[\protect\citeauthoryear{VanderPlas \& Ivezic}{VanderPlas \&
  Ivezic}{2015}]{VanderPlas2015}
VanderPlas J.~T.,  Ivezic Z.,  2015, \mn@doi [The Astrophysical Journal]
  {10.1088/0004-637X/812/1/18}, 812, 18

\bibitem[\protect\citeauthoryear{Vincent et~al.,}{Vincent
  et~al.}{2017}]{Vincent2017}
Vincent J.~B.,  et~al., 2017, \mn@doi [Monthly Notices of the Royal
  Astronomical Society] {10.1093/mnras/stx1691}, 469, S329

\bibitem[\protect\citeauthoryear{Weissman, Doressoundiram, Hicks, Chamberlin,
  Sykes, Larson  \& Hergenrother}{Weissman et~al.}{1999}]{Weissman1999}
Weissman P.~R.,  Doressoundiram A.,  Hicks M.~D.,  Chamberlin A.,  Sykes M.~V.,
   Larson S.,   Hergenrother C.,  1999, Bulletin of the Astronomical Society,
  31, 1121

\makeatother
\end{thebibliography}

\bsp	
\label{lastpage}
\end{document}